\documentclass[aip,
amsmath,
amssymb,
reprint]{revtex4-1}

\usepackage{graphicx}
\usepackage{dcolumn}
\usepackage{bm}

\usepackage[utf8]{inputenc}
\usepackage[T1]{fontenc}
\usepackage{mathptmx}
\usepackage{etoolbox}
\usepackage{amsthm}
\newtheorem{theorem}{Theorem}
\usepackage{hyperref}
\usepackage{float}
\usepackage{tikz}
\usetikzlibrary{shapes.geometric, arrows}
\usepackage{physics}
\usepackage{subcaption}

\DeclareUnicodeCharacter{2212}{-}

\makeatletter
\def\@email#1#2{%
 \endgroup
 \patchcmd{\titleblock@produce}
  {\frontmatter@RRAPformat}
  {\frontmatter@RRAPformat{\produce@RRAP{*#1\href{mailto:#2}{#2}}}\frontmatter@RRAPformat}
  {}{}
}%
\makeatother
\begin{document}

\preprint{AIP/123-QED}





\title[Effect of clustering on Turing instability in complex networks]{Effect of clustering on Turing instability in complex networks}
\author{Samana Pranesh}
\affiliation{ 
The Uncertainty Lab, Department of Applied Mechanics \& Biomedical Engineering, Indian Institute of Technology Madras, Chennai - 600036, Tamil Nadu, India
}%
\author{Devanand Jaiswal}
\affiliation{ 
Department of Chemistry, Saint Louis University, Saint Louis, MO, USA
}%
\author{Sayan Gupta}%
 \email{sayan@iitm.ac.in}
\affiliation{ 
The Uncertainty Lab, Department of Applied Mechanics \& Biomedical Engineering, Indian Institute of Technology Madras, Chennai - 600036, Tamil Nadu, India
}%

\affiliation{%
Complex Systems and Dynamics Group, Indian Institute of Technology Madras, Chennai - 600036, Tamil Nadu, India
}%

\date{\today}

\begin{abstract}
Turing instability in complex networks have been shown in the literature to be dominated by the distribution of the nodal degrees. {The conditions for Turing instability have been derived with an explicit dependence on the eigenvalues of the Laplacian, which in turn depends on the network topology. This study reveals that apart from average degree of the network, another global network measure - the nodal clustering - also plays a crucial role.} Analytical and numerical results are presented to show the importance of clustering for several network topologies ranging from the  $\mathbb{S}^1$ / $\mathbb{H}^2$ hyperbolic geometric networks that enable modelling the naturally occurring clustering in real world networks, as well as the random and scale free networks, which are obtained as limiting cases of the $\mathbb{S}^1$ / $\mathbb{H}^2$ model. Analysis of eigenvector localization properties in these networks are shown to reveal distinct signatures that enable identifying the so called Turing patterns even in complex networks.

\end{abstract}

\maketitle
\begin{quotation}
The competing dynamics of multiple diffusive agents that behave nonlinearly in spatio-temporal systems, under certain conditions, lead to Turing instability that manifests as patterns. The corresponding weak form representation of the underlying partial differential equations, obtained through discretization of the spatial domain,  lead to a system of dynamical units (nodes) that are in the form of a regular lattice network. Generalization of these results to complex networks has shown the possibility of Turing instability in such network topologies as well. {Turing instability, being a diffusion-induced process, is primarily influenced by the nodal degrees. Consequently, it can be triggered in any network by tuning the average degree, as established in the literature.} This study however shows that network clustering can induce Turing instability even for networks whose average network degree do not fulfill the conditions for Turing instability.  The results presented are shown to be valid for a suite of complex network models, ranging from the $\mathbb{S}^1$ / $\mathbb{H}^2$ hyperbolic geometric network model as well as its limiting cases, the random and scale free networks.  Though the manifestation of Turing instability as patterns are not easily discernible in complex network topologies, the analysis of eigenvector localization properties are shown to reveal distinct signatures that enable identifying Turing instability even in complex networks.  

\end{quotation}

\section{Introduction}
The interaction between two or more competing diffusive agents in a medium usually results in a uniform homogenized stable state. However, under certain conditions depending on the difference in the diffusivity between the competing agents, there is a loss of stability of the uniform homogenized states leading to 
repetitive concentrations of these agents spatially that manifest as stable temporal patterns, known as Turing patterns \cite{turing1952chemical}. The formation of these patterns is the outcome of non-equilibrium self organization following Turing instability of the homogenized state, and has been widely observed in a host of dynamical systems
\cite{cross1993pattern,cross2009pattern,gierer1972theory,koch1994biological,kondo2010reaction,meinhardt1982models,walgraef2012spatio,zhabotinsky1995pattern}. In its simplest form, Turing instability has been mathematically investigated with two competing agents, one  
being the activator and the other being the inhibitor and is represented by a set of coupled partial differential equations (referred to as reaction-diffusion (RD) equations) of the form \cite{prigogine1968symmetry}
\begin{align}
\frac{ \partial u}{\partial t} &= f(u,v) + D_{u}\nabla^2 u, \label{int1}\\
\frac{\partial v}{\partial t} &= g(u,v) + D_{v}\nabla^2 v,
\label{int2}
\end{align} 
where, the state variables $u\equiv u({\bf x},t)$ and $v \equiv v({\bf x},t)$ are the local densities of two distinct competing agents 
and are  functions of the spatial variable ${\bf x}$ and the temporal variable $t$, $f(u,v)$ and $g(u,v)$ are generally nonlinear functions that govern the temporal evolution of   $u$ and $v$ respectively, $D_u$, $D_v$ are the corresponding diffusion coefficients and $\nabla^2(\cdot)$ is the Laplace operator.   Typically, one of the state variables, say $u$, promotes the growth of both state variables and is identified as the activator, while the other state variable, $v$, plays the  role of an inhibitor. If the diffusion of the inhibitor  is faster than the activator, sharp regions of concentration differences are observed for the active agent leading to stable Turing patterns.  The  formation and geometry of Turing patterns depend on the functions $f(\cdot)$, $g(\cdot)$, as well as on the inequality in the diffusion coefficients,  defined in terms of the normalized coefficient $\sigma = D_v/D_u$. The pattern formation is  an outcome of diffusion induced instability of the stable homogenized state (when there is no diffusion), leading to temporally stable but spatially inhomogeneous solutions. 
Diffusion in the continuous media is modelled by Brownian particles, and by random walk, when the RD equations are discretized in a regular lattice. In the latter, all the nodes have equal degrees and is therefore a homogeneous network.
However, Turing instability is not limited to only homogeneous networks but have been observed in complex networked systems as well \cite{othmer1971instability,horsthemke2004network,moore2005localized,asllani2014theory,muolo2019patterns,petit2017theory,van2021theory,gallo2022synchronization,gao2023turing,asllani2014turing,asllani2015turing,kouvaris2015pattern,busiello2018homogeneous,siebert2020role}. As is well known, complex networks are heterogeneous networks with the nodal degrees following a distribution. Diffusion in such networks follow a different process and has been shown to be appropriately modelled by continuous time random walk, where the jump rates at each node is proportional to its degree\cite{van2023emergence}. The diffusive particles of the two competing species further interact with each other nonlinearly, leading to temporal fluctuations in the concentration at each node.    Assuming that the number of diffusive particles to be large and focusing on the average concentrations $u_i(t)$ and $v_i(t)$ of the two competing agents at each node, the corresponding evolution equations can be expressed as  
\begin{align}
\dot{\bf u} & = f({\bf u},{\bf v}) + D_u {\bf L}{\bf u},\label{int1a}\\
\dot{\bf v} & = g({\bf u},{\bf v}) + \sigma D_u {\bf L} {\bf v}.\label{int2a}
\end{align}
%
%
%
Here,  ${\bf u}$, ${\bf v}$ are vectors representing the average concentrations at the nodes and the Laplacian operator in Eqs.(\ref{int1})-(\ref{int2}) is replaced by the Laplacian matrix ${\bf L}$, defined as $L_{ij}=A_{ij}-k_i\delta_{ij}$\cite{van2023graph}, with $A_{ij}$ being the symmetric adjacency matrix with elements $1$ if there is a connection with nodes $i$ and $j$ and $0$ otherwise. The degree of node $i$ is given by $k_i=\sum_{j=1} ^N A_{ij}$ and the flux into node $i$ is $\sum_{j=1}^N L_{ij} u_j$.  Here, $N$ is the total number of nodes in the network. 
At the domain boundaries, either a zero flux or periodic boundary conditions are imposed.  The conditions for Turing instability have been analytically shown to be dependent on the average degree ${\rm E}[k]$ \cite{nakao2010turing}. This is not surprising as the diffusion rates at each node is proportional to the degree $k$ of the node and the average degree ${\rm E}[k]$ dictates the mean field behaviour. 
Here, ${\rm E}[\cdot]$ is the expectation operator.

While Turing instability is accompanied with the formation of Turing patterns in continuous media or degree regular lattices, no such geometric patterns are usually discernible in complex networks, though hints of patterns have been observed in simple network models \cite{asllani2014turing,asllani2016tune,cencetti2020generalized,hutt2022predictable}. The lack of observable patterns, which are essentially spatial structures,  can be attributed to the nodes in complex networks not having any spatial coordinate attributes and being characterized only in terms of node connectivity for which there are no unique metrics. 
However, recent studies on network geometry \cite{boguna2021network} have shown that real world complex networks are associated with a latent metric space with hyperbolic geometry \cite{krioukov2010hyperbolic}, which enables simple explanations for observed topological features, such as heterogeneous distributions \cite{serrano2008self}, clustering \cite{candellero2016clustering, fountoulakis2021clustering},  small-worldness \cite{abdullah2015typical}, spectral, and  self-similarity. This has led to the development of the geometric soft configuration model, also referred to as the  $\mathbb{S}^1$ / $\mathbb{H}^2$ network model \cite{krioukov2010hyperbolic,serrano2008self,serrano2021shortest}. This model  combines a similarity dimension encoded in a one dimensional sphere and a popularity dimension based on the degree of each node. The  nodes are embedded in the  metric space, and the connection probability depends on the distance among the nodes in the metric space. Importantly, the geometric soft configuration model is able to capture the properties of real world networks, such as  degree  heterogeneity, high-clustering, small-world properties and community structure\cite{garcia2018geometric}. The Erdos-Renyi (ER) random network \cite{erdds1959random} and the Barabasi-Albert scale free network \cite{barabasi1999emergence} are shown to be limiting cases. A recent study on Turing instability \cite{van2023emergence}  within the framework of such geometric random graph models have shown that geometric patterns are clearly discernible in the underlying geometric space, and have quantified these patterns in terms of their wavelengths.

For reasons that have already been discussed,  investigations on Turing instability in complex networked systems have been predominantly carried out with 
${\rm E}[k]$ as the primary control parameter \cite{nakao2010turing, guo2021turing,liu2020turing}. These investigations focussed on investigating the conditions leading to Turing instability in complex networks, with the numerical investigations being carried out on   synthetic networks, such as the ER or the scale free networks. The topology of these networks and hence  ${\rm E}[k]$, are easily controllable with parameters of these models. However, studies have shown that in addition to degree heterogeneity, clustering is also a crucial network parameter and one can have two networks with the same degree distribution but very different clustering characteristics. In fact, while ER and scale free networks are known to have low clustering, many real world networks have community structures with the formation of clusters of nodes and have high clustering\cite{nr}; see Table \ref{tab:networks}. 
%
%
\begin{table}[htbp]
\begin{tabular}{lllll}
\hline
\multicolumn{1}{c}{\textbf{Network}} & \multicolumn{1}{c}{\textbf{Nodes}} & \multicolumn{1}{c}{\textbf{Edges}} & \multicolumn{1}{c}{\textbf{E[k]}} & \multicolumn{1}{c}{\textbf{C}} \\ \hline 
Citation                               & 22.9K                               & 2.7M                                & 233                                          & 0.809                                                 \\ 
\textit{C.elegans}                     & 453                                 & 2K                                  & 8                                            & 0.6464                                                \\ 
Yeast protein interaction              & 6K                                  & 313.9K                              & 104                                          & 0.1636                                                \\ 
USAir97 Infrastructure                 & 332                                 & 2.1K                                & 12                                           & 0.625                                                 \\ 
Dublin human contact           & 410                                 & 2.8K                                & 13                                           & 0.558                                                 \\ \hline
\end{tabular}
\caption{Topological properties of some empirical networks; $C$ is the clustering co-efficient}
\label{tab:networks}
\end{table}
This seems to indicate that clustering could also play an important role in Turing instability in complex networks, an aspect which does not seem to have been studied in the literature.
This study is aimed towards addressing this lacuna. 
%
%
%
{The manuscript is organised as follows: Section \ref{background} provides a brief description of Turing instability in complex networks. Section \ref{section3} provides a background on geometric networks and the properties of Laplacian spectrum is discussed in Section \ref{mm}.
The problem statement is outlined in Section \ref{results}. Results and related discussions are presented in Section \ref{discussion}. The salient outcomes of this study are summarized in the
concluding section in Section \ref{conclusion}}.

\section{Review of Turing instability in complex networks}
\label{background}

For the system of RD equations in Eqs.(\ref{int1a})-(\ref{int2a}),  let $(u_i^*,v_i^*)$  be the positive uniform stationary state for  $f(u,v)$ and $g(u,v)$  $\forall \,\, i$, when $D_u=0$ (indicating absence of diffusion). Therefore, $(u_i^*,v_i^*)$ are obtained as the solution of the coupled set of equations $f(u_i,v_i)=0$ and $g(u_i,v_i)=0$. For ease of notation, the subscripts $i$ are dropped henceforth. The stationary solution $(u^*,v^*)$ are linearly stable in the absence of diffusion, if
\begin{eqnarray}
f_u>0,\,\,  f_v <0,\,\,  g_u>0,\,\,  g_v<0,
\end{eqnarray}
\begin{eqnarray}
(f_u +g_v) < 0 ,\,\,\, f_ug_v -f_vg_u >0,
\end{eqnarray}
where, $f_u,f_v, g_u, g_v$ denote the partial derivatives of $f(\cdot)$ and $g(\cdot)$ with respect to $u$ and $v$ and evaluated at $(u^*,v^*)$; see \cite{nakao2010turing} for  details.
As ${\bf L}$ is symmetric, real and negative semi-definite, its eigenvalues $\alpha_s \leq 0$, ($s=1,\hdots, N$) and the corresponding eigenvectors ${\boldsymbol \phi}^{s} = [{\boldsymbol \phi}_1^{s},\hdots,{\boldsymbol \phi}_N^{s}]$ span a $N$-dimensional basis. Note that $\sum_{j=1}^{n}L_{ij} {\boldsymbol \phi}_j^s=\alpha_s{\boldsymbol \phi}_i^s$. 
Let $\Delta u_i$ and $\Delta v_i$ be  small perturbations to $(u^*, v^*)$. Substituting  in Eqs.\eqref{int1}-\eqref{int2} and applying Taylor series expansion upto first order terms, it can be shown that 
\begin{align}
\frac{{\rm d}}{{\rm d}t} \Delta u_{i} &=f_u(u^*,v^*)\Delta u_i+f_v(u^*,v^*)\Delta v_i
+D\sum_{j=1}^{N}l_{ij}\Delta u_{j},\label{net_m11}\\
\frac{{\rm d}}{{\rm d}t}\Delta v_{i}&=g_u(u^*,v^*)\Delta u_i+g_v(u^*,v^*)\Delta v_i
+\sigma D\sum_{j=1}^{N}l_{ij}\Delta v_{j}.\label{net_m22}
\end{align}
Expressing the perturbations as linear combination of the eigenvectors as
\begin{align}
\Delta u_i(t)&=\sum_{s=1}^{N}A_s \exp(\gamma_s t){\boldsymbol \phi}_i^s, \label{pert1}\\
\Delta v_i(t)&=\sum_{s=1}^{N}A_sB_s \exp(\gamma_s t){\boldsymbol \phi}_i^s,\label{pert2}
\end{align}
 substituting in Eqs.\eqref{net_m11}-\eqref{net_m22} and using the orthogonality conditions, lead to the coupled equations
\begin{align}
\gamma_s\begin{pmatrix}
1\\
B_s
\end{pmatrix}
=
\begin{pmatrix}
f_u+D\alpha_s & f_v\\
g_u & g_v+D\sigma \alpha_s
\end{pmatrix}
\begin{pmatrix}
1\\
B_s
\end{pmatrix}
\label{eq:ev}
\end{align}
for  $s=1,\hdots, N$. Eq.\eqref{eq:ev} represents an eigenvalue problem with $\alpha_s$ being the eigenvalue of the Laplacian, the characteristic equation for which can be written as
\begin{align}
\gamma_s^2 + b(\alpha_s)\gamma_s + c(\alpha_s) = 0.
\label{char_eq}
\end{align}
Here, $b(\alpha_s) = -[Tr({\bf J}) + D(1+\sigma)\alpha_s]$ and $c(\alpha_s) = [\sigma D\alpha_s^2 + D(g_v + \sigma f_u)\alpha_s + |J|]$, ${\bf J} = (\begin{matrix}f_u & f_v\\ g_u & g_v\end{matrix})$, with $Tr({\bf J}) = f_u+g_v$ and $|{\bf J}|=f_ug_v-f_vg_u$.
Solving Eq.\eqref{char_eq} for $\gamma_s$ leads to
\begin{eqnarray} \label{gamma}
\gamma_s &= &\frac{1}{2}[ f_u+g_v+(1+\sigma)D \alpha_s \nonumber \\
& & \pm \sqrt{4f_vg_u+\left(f_u-g_v+(1-\sigma)D\alpha_s\right)^2}].\label{root}
\end{eqnarray}
$\gamma_s$ represents the growth rate of the perturbations and as can be seen, depends explicitly on the diffusivity $D\alpha_s$.  As Turing instability is induced by growing perturbations, {\it i.e.},  when  $\gamma_s>0$, it follows that the criterion for instability can be given by $\Re(\gamma_s)>0$, where one considers the larger of the two roots. A bifurcation diagram can be obtained for $\gamma_s$ with $\alpha_s$ being the bifurcation parameter to identify the regime $\Re(\gamma_s)>0$. The critical value of $\sigma$  for which $\Re(\gamma_s)<0$ can be shown to be given by
\begin{align}
\sigma_c=\left[f_vg_v-2f_vg_u+2\left[f_vg_u(f_vg_u-f_ug_v)\right]^{1/2}\right]/f_u^2, \label{sigm_th}
\end{align}
where $\sigma_c$ is the critical value.   
It can be further  shown that for $\Re(\gamma_s)>0$,  the condition $c(\alpha_s) < 0$ should be satisfied. In other words, 
\begin{align}
    c(\alpha_s) = [\sigma D\alpha_s^2 + D(g_v + \sigma f_u)\alpha_s + |{\bf J}|] < 0.
    \label{eqn_c}
\end{align}
The expression for $c(\alpha_s)$ in Eq.\eqref{eqn_c} represents an upward opening parabola, since $\sigma D\alpha_s^2 > 0$. The roots of $c(\alpha_s)=0$ are given by 
\begin{align}
    \alpha_{s_{1,2}} = \frac{1}{2\sigma D}\Big{(}-[g_v + g_u\sigma] \pm  \sqrt{(g_v + \sigma f_u)^2 - 4\sigma |{\bf J}|} \Big{)}. 
   \label{eigen}
\end{align}
It follows that $c(\alpha_s)<0$  if some of the Laplacian eigenvalues, $\alpha_s$, fall within the range $[\alpha_{s_1},\alpha_{s_2}]$. 
It has been shown in \cite{nakao2010turing} that $\Re(\gamma_s)$ attains its maximum positive value for
\begin{align}
    \alpha_{max} = \frac{1}{D(1-\sigma)}\Big{[}f_u -g_v - (1+\sigma)\sqrt{\frac{f_vg_u}{\sigma}} \Big{]}.
     \label{eigen_max}
\end{align}
As $Tr({\bf L}) = \sum_i k_i = \sum_s \alpha_s$,  it follows that the average nodal degree in a network  is also equal to the average values of $\alpha_s$, {\it i.e.}, ${\rm E}[k] = {\rm E} [\alpha]$.  The implication here is that for a given set of diffusion constants and system parameters,  Turing instability is dictated by the eigenspectrum of the Laplacian matrix ${\bf L}$. These conditions are general and are independent of the forms of $f(\cdot)$ and $g(\cdot)$ and therefore has been observed in a wide variety of systems ranging from chemical reactions\cite{prigogine1968symmetry,castets1990experimental,ouyang1991transition},  biological morphogenesis\cite{harris2005molecular} to ecosystems\cite{segel1976application,mimura1978diffusive,maron1997spatial,gibert2019laplacian}. Most of the studies in the literature use  the global network topological parameter ${\rm E}[k]$ as the control parameter to investigate the Turing instability regime.


\section{
Review of  $\mathbb{S}^1$ / $\mathbb{H}^2$ network model}\label{section3}
Many real-world networks across various domains exhibit notable similarities, including scale-free degree distribution, sparseness, high clustering, small-world properties, and community structure \cite{garcia2018geometric}. The degree distribution 
 in the commonly used Erd{\~o}s-R{\'e}nyi (ER) random network \cite{erdds1959random} model follows a Poisson distribution, ruling out the occurrence of hubs and leading to clustering  that diminishes as $N \rightarrow \infty$. The Barabási-Albert model \cite{barabasi1999emergence} for the scale free network enables presence of hubs through its preferential attachment to nodes with high degrees, leading to a power law degree distribution. However, here too the clustering diminishes as $N \rightarrow \infty$, indicating that while node popularity is influential in network formation, it is not the only governing parameter. The small world Watts-Strogatz network \cite{watts1998collective} model lead to networks with high clustering but a degree distribution that is almost homogeneous.

The $\mathbb{S}^1$ / $\mathbb{H}^2$ network is an alternative model that allows separately tuning the clustering parameter in the network using the concept of node similarity \cite{boguna2021network,krioukov2010hyperbolic}. 
The underlying principle here is that nodes possess intrinsic properties beyond popularity, and the similarity between these properties influences the likelihood of interaction between two nodes. For example, if nodes $A$ and $C$ share similarity with a third node $B$, it is plausible that $A$ and $C$  also exhibit similarity. 
%
%
As has been mentioned, the $\mathbb{S}^1$ / $\mathbb{H}^2$ network is associated with a latent metric space, where  a similarity dimension is
encoded in a one dimensional sphere and a popularity dimension based on the degree of each node. Each node can therefore be assumed to be associated with a pair of hidden variables $\kappa \in [\kappa_0,\infty)$ and $\theta_i \in [0,2\pi)$. $\kappa$ and $\theta$ are statistically independent random variables where the former accounts for the ensemble average degree  of the $i$-th node,  which can be generated from an arbitrary distribution $\rho(\kappa)$ and the latter is the angular position of node  $i$ on the circle which is homogeneously distributed. The distance between two points in this space serves as a measure of their dissimilarity.
%
Therefore, clustering is naturally embedded within the structure of the metric space. 
%
The degree distribution is assumed to follow a power-law distribution of the form 
\begin{equation}
\rho(\kappa) = (\Lambda-1)\kappa_0^{\Lambda-1}\kappa^{-\Lambda},\,\, \kappa \geq \kappa_0.
\end{equation}
Degree heterogeneity is modulated by the power-law exponent $\Lambda$ and the limit $\Lambda \rightarrow \infty$ is equivalent to a homogeneous degree distribution of hidden degrees resulting in ER network \cite{krioukov2010hyperbolic}. The  connection probability between nodes $i$ and $j$ is given by
\begin{align}
    p_{ij}=\frac{1}{1+(\frac{x_{ij}}{\mu \kappa_i \kappa_j})^\beta},
    \label{eq:connection}
\end{align}
where, $x_{ij}$ is the distance between the nodes in a circle of radius $R=N/2\pi$, $\beta>1$ is a parameter which controls the clustering in the network 
and $\mu$ is a parameter that controls the average degree $E[k]$ of the network. 

Interestingly, it was discovered that the hyperbolic geometry - metric space of constant negative curvature - emerges from the $\mathbb{S}^1$ model. The hidden degree $\kappa$, associated with each node,  is mapped into a radial co-ordinate through the transformation
\begin{align}
    r = R_{\mathbb{H}^2} - 2 \ln \frac{\kappa}{\kappa_0}.
\end{align}
This mapping pushes the higher-degree nodes closer to the origin of the co-ordinate system 
and the connection probability in Eq.\eqref{eq:connection} can be rewritten as
\begin{align}
    p_{ij} = \frac{1}{1+e^{(\beta/2)(d_{\mathbb{H}^2, ij} - R_{\mathbb{H}^2})}},
\end{align}
where, $R_{\mathbb{H}^2} = 2 \ln(N/\mu\pi\kappa_0^2)$ is the radius of the hyperbolic disk and $d_{\mathbb{H}^2, ij}$ is the hyperbolic distance between nodes $i$ and $j$ given by the hyperbolic law of cosines:
\begin{align}
    \cosh(d_{\mathbb{H}^2, ij}) = \cosh r_i \cosh r_j - \sinh r_i \sinh r_j \cos \Delta\theta_{ij}.
\end{align}
Here, $\Delta\theta_{ij}$ is the angular separation between the nodes. The $\mathbb{S}^1$ model is therefore isomorphic to a purely geometric model, the $\mathbb{H}^2$ model,  in which nodes are distributed on a hyperbolic disk and then connected with a probability that decreases with the distance among every pair. For more details, see \cite{boguna2021network,krioukov2010hyperbolic}. 

\section{Clustering and the Laplacian spectrum}\label{mm}

The focus of this study is to investigate the effects of clustering  on Turing instability in complex networks, having different classes of network topologies. The effects of clustering in synchronization in complex networks is well documented \cite{mcgraw2007analysis}. However, to the best of the knowledge of the authors, its effect on Turing instability has not been investigated. The underlying objective behind this study is  the premise that clustering is an additional parameter that affects Turing instability, independently of the global network parameter ${\rm E}[k]$ that has been traditionally used to characterize the Turing instability regimes in dynamical systems associated with complex networks.

Clustering, also known as transitivity, refers to the tendency of two nodes which share a common neighbour, to have an increased likelihood of also being directly connected to each other \cite{watts1998collective}. This therefore  also indicates the presence of triangles in the  network. The degree of clustering in a network is quantified through the clustering co-efficient ($C$), 
defined as an average of local clustering co-efficient associated with each node $i$. Mathematically,  $C$ is defined as \cite{newman2018networks},
\begin{align}
   C= \frac{1}{N} \sum_{i=1} ^N T_i,\,\,\,\,\,\, T_i = \frac{{2 \cdot e_i}}{{k_i \cdot (k_i - 1)}},
\end{align}\label{cc}
where 
$e_i$ is the number of edges between the neighbors of node $i$ and $k_i$ is the degree of node $i$. 

\begin{figure*}[t]
\centering
\includegraphics[width=0.8\textwidth]{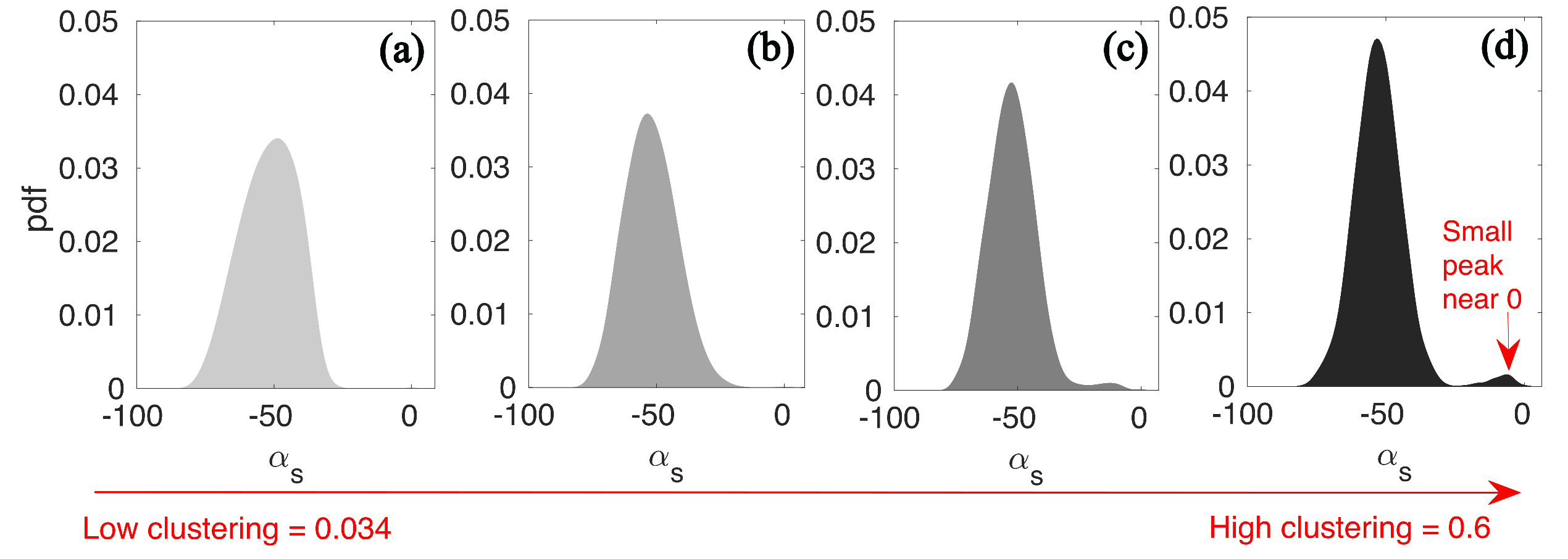}
    \caption{Eigenvalue distribution of the Laplacian for an ER network with $E[k]=44$ for increasing value of clustering co-efficient $C$.}
\label{fig:diff_cc_spectra_ER}
\end{figure*}

To investigate the independent effects of clustering on Turing instability in complex networks, it is essential to generate networks whose clustering coefficient can be varied without changing ${\rm E}[k]$. This is is achieved through the ``ClustRNet" algorithm \cite{bansal2009exploring},  that enables  generating networks with a predetermined clustering coefficient while keeping the degree distribution unchanged. This holds true for geometric networks as well.
%
%
Figure \ref{fig:diff_cc_spectra_ER} shows the variation of the  spectra of the eigenvalues of the Laplacian for a ER network with ``natural" low clustering (left most) with increasing value of $C$.  
The degree distribution associated with all the four cases shown in Fig.\ref{fig:diff_cc_spectra_ER} are identical, with $E[k]=44$. It is observed that increasing the clustering modifies the shape of the Laplacian eigenvalue spectrum, causing the peak to sharpen. {
For ER networks, the shape of the eigenspectrum is Poisson-like single peak centered around $E[k]$ and appears to follow the shape of the degree distribution.} 
 As the clustering coefficient increases, the cluster of eigenvalues near zero separates from the main peak and a secondary peak is observed close to zero; see Fig. \ref{fig:diff_cc_spectra_ER}(c)-(d). 
 
 {The appearance of a peak close to zero in the eigenvalue spectrum for high values of $C$ can be  mathematically explained.
It is  known that the Laplacian matrix $\mathbf{L}$ of a network always has a zero eigenvalue. To prove this,  consider the all-ones vector \( \mathbf{1} \). Multiplying the Laplacian matrix \( \mathbf{L} \) by this vector, it can be shown that 
    \begin{equation}
    \mathbf{L} \cdot \mathbf{1} = (\mathbf{D} - \mathbf{A}) \cdot \mathbf{1} = \mathbf{d} - \mathbf{d} = \mathbf{0}.
    \end{equation}
    Here \(\mathbf{D} \) is the degree matrix and \( \mathbf{A} \) is the adjacency matrix. The above result follows as \( \mathbf{A} \cdot \mathbf{1} \) yields the degree vector \( \mathbf{d} \), and \( \mathbf{D} \cdot \mathbf{1} \) is also \( \mathbf{d} \).
Therefore, \( \mathbf{1} \) is an eigenvector corresponding to the eigenvalue 0. 

It is also known that the number of zero eigenvalues of ${\bf L}$ is equal to the number of connected components in the network. For a network \( G \) with \( k \) connected components, the Laplacian matrix can be block-diagonalized as:
    \[
    \mathbf{L} = \begin{pmatrix}
    {\bf L}_1 & 0 & \cdots & 0 \\
    0 & {\bf L}_2 & \cdots & 0 \\
    \vdots & \vdots & \ddots & \vdots \\
    0 & 0 & \cdots & {\bf L}_k
    \end{pmatrix}
    \]
    where each \( {\bf L}_i \) is the Laplacian matrix of the \( i \)-th connected component. Each \( {\bf L}_i \) has exactly one zero eigenvalue because each connected component is represented by its own all-ones vector ${\bf 1}_i$. Thus, the total number of zero eigenvalues of \( \mathbf{L} \) equals the number of connected components in the network.

 The presence of eigenvalues very close to zero can indicate bottlenecks in the network. This can be explained using the Cheeger's inequality\cite{van2023emergence,polya1951isoperimetric,cheeger1970lower}, which states that 
    \begin{align}\label{cheeger}
    \frac{\alpha_2}{2} \leq h(G) \leq \sqrt{2\alpha_2 \Delta},
    \end{align}
    where \( \alpha_2 \) is the second smallest eigenvalue of the Laplacian (also known as algebraic connectivity), \( \Delta \) is the maximum degree of the network and \( h(G) \) is the Cheeger constant. The Cheeger constant $h(G)$
is a measure of the well-connectedness the network $G$  and is defined as
\begin{align}
    h(G) = \min_{S \subset V, 0 < |S| \leq |V|/2} \frac{|S|}{|\partial S|}.
\end{align}
Here, $V$ is the set of vertices in the network, $S$ is a subset of 
$V$ and $\partial S$  is the number of edges that have one endpoint in 
$S$ and  the other endpoint in $V\setminus S$. For small values of \( \alpha_2 \), it follows from Eq.\eqref{cheeger} that \( h(G) \) must also be small. A small value of Cheeger constant implies the existence of a  subset \( S \) with relatively few edges linking it to its complement \( V \setminus S \). This suggests either poor overall network connectivity or the presence of loosely interconnected regions. Hence, eigenvalues close to zero indicates the presence of robust communities or nearly disjointed components within the network. These strong communities typically feature sparse connections between groups compared to within groups, resembling nearly disjointed components, thereby yielding small but nonzero eigenvalues \cite{mcgraw2008laplacian, mohar1991laplacian}.}


\section{Problem statement} \label{results}

As has been shown in Sec. \ref{background}, the analytical conditions for Turing instability for a general RD system depends on the critical value of the diffusivity ratio $\sigma_c$ and is independent of the form of  $f(\cdot)$ and $g(\cdot)$. A parametric analysis enables identifying the regime of Turing instability, by identifying the regime in which  $\Re(\gamma_s) >0$. The independent effect of clustering can be demonstrated by inducing Turing instability in RD systems where $C$ is used as the control parameter, keeping the degree distribution and hence ${\rm E}[k]$, unchanged. The effects of $C$ on Turing instability are numerically investigated in this study, along with explanations on the underlying physics.

Since numerical results can be presented  for only specific functional forms of $f(\cdot)$ and $g(\cdot)$,
%
%
for the sake of illustration, in this study these are taken to be of the form
\begin{align}
f(u_i,v_i) &=u_i(1-u_i)-\frac{u_i^2v_i}{au_i^2+1}, \label{med1}\\
g(u_i,v_i) &=bv_i(1-\frac{cv_i}{u_i}). \label{med2}
\end{align}  
The corresponding RD equations represent a well established prey-predator ecological system, with $u_i$ and $v_i$ representing densities  of the prey and predator species at node $i$ and $a$, $b$, and $c$ are strictly positive (real) constants. It can be seen that both $u_i$ and $v_i$ follow a logistic growth model. The reduction in prey population exhibits the functional response of Holling type III,  while the reduction of predators follows the Leslie-Gower functional response\cite{liu2020turing}. In the context of this ecological system, $D_u$ and $D_v$ in Eqs.\eqref{int1}-\eqref{int2} are taken to be $D$ and $\sigma D$, representing the prey and predator diffusion mobilities. The numerical values of the parameters are taken to be $a=80$, $b=0.6$, $c=0.016$ and $D=0.01$ unless mentioned otherwise. In the absence of diffusion, the equilibrium state is $u^*=0.30$ and $v^*=u^*/c$. The initial conditions are considered as  small perturbations of the homogeneous state $(u^*,v^*)$.  The critical value of $\sigma_c$ is calculated from Eq.\eqref{sigm_th} and is  $25.049$.  {In the following section, the effect of clustering on  is investigated for (a) Erdos-Reyni (ER) random network with natural low clustering (b) ER network with high clustering (c) scale free network and (d) $\mathbb{S}^1$ / $\mathbb{H}^2$ network models.}

\section{Numerical Results and discussions}\label{discussion}

\subsection{ER network with natural low clustering}

\begin{figure}[htbp]
\centering
\includegraphics[width=0.5\textwidth]{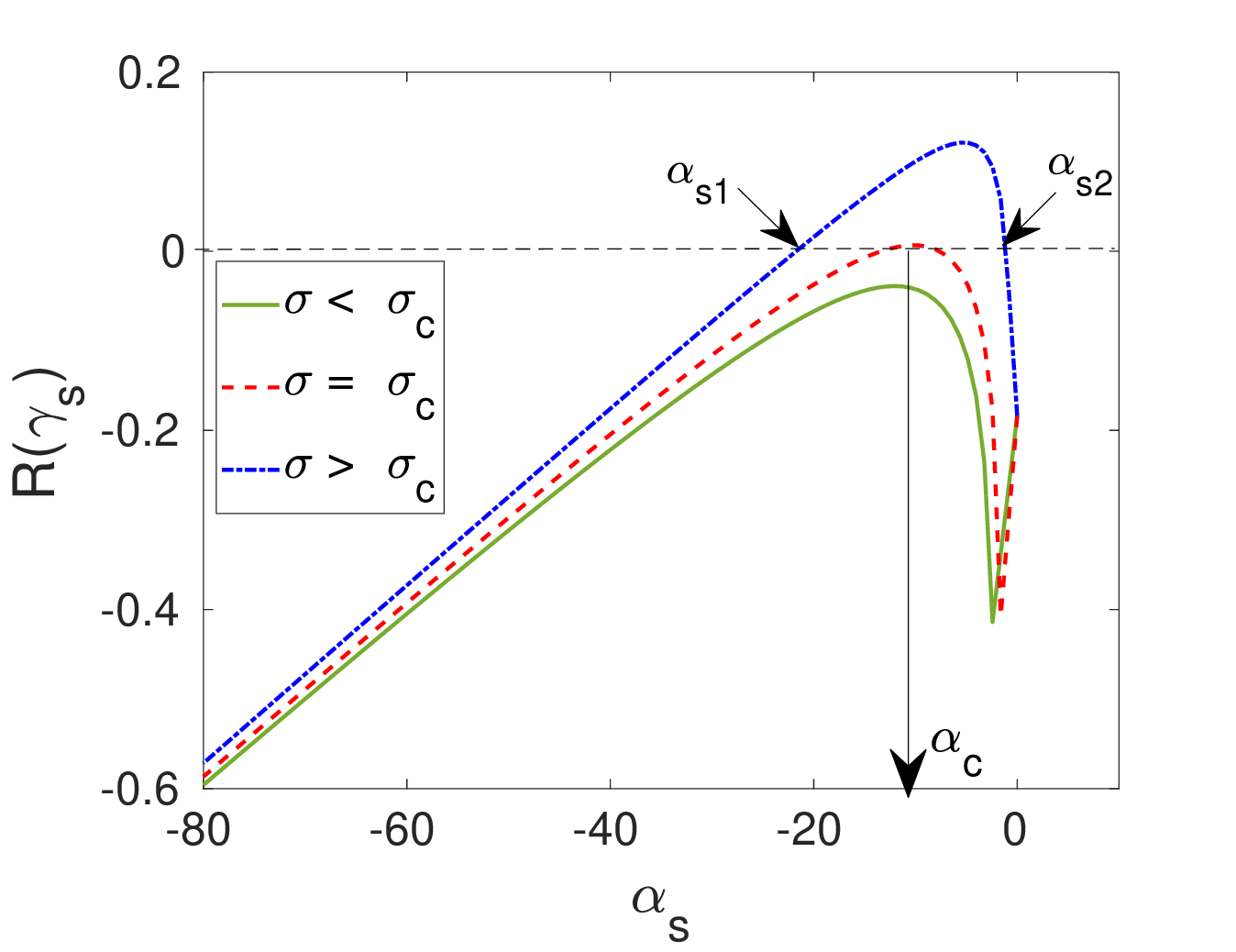}
    \caption{Variation of $\Re(\gamma_s)$ with $\alpha_s$; blue, red and green lines represent Eq.\eqref{gamma} for the cases $\sigma>\sigma_c$, $\sigma=\sigma_c$ and $\sigma<\sigma_c$ respectively.}
    \label{fig:eigendist1}
\end{figure}

ER networks \textcolor{black}{comprising of $N=1225$ nodes is constructed, with ${\rm E}[k]$ being  one of the control parameter that is varied from $4$ to $44$. The corresponding $C$ for these ER networks are shown to vary from $0.0043$ and $0.0341$ and confirms that clustering is naturally low for finite sized ER networks. As discussed in Section \ref{background},  by solving Eq.\eqref{char_eq} for $\gamma_s$ and identifying the modes where $\Re(\gamma_s)>0$, the regime for Turing instability for the RD  system can be identified.} 
As shown in Eq.\eqref{root}, $\gamma_s$ depends on the eigenvalues $\alpha_s$ of the Laplacian matrix ${\bf L}$ and  therefore is implicitly dependent on the eigenspectrum associated with the network topology. The dependence on   the diffusivity ratio $\sigma$ is explicit. {Fig. \ref{fig:eigendist1} shows the variation of $\Re(\gamma_s)$ as a function of $\alpha_s$ for this RD problem, for three different cases of $\sigma$.}
It is observed that the curve corresponding to $\sigma<\sigma_c$ always lies below zero, indicating that there is no possibility of Turing Instability. {For $\sigma = \sigma_c$, the peak of the curve is tangent to the zero line and the corresponding  value of $\alpha_s$ is defined as the critical value $\alpha_c$; see Fig.\ref{fig:eigendist1}. The corresponding Laplacian mode $\boldsymbol{\phi^{\alpha_c}}$ is defined as  the critical mode. The implication for the critical case of $\sigma=\sigma_c$ is that the perturbations corresponding to all the Laplacian modes die down at large time $t$ except for the critical mode. From Eq.\eqref{pert1}, it follows that the instantaneous evolution can be expressed as 
\begin{align}
\label{perturb1}
u_i(t) = u^* + A_c \exp(\gamma_c t) \boldsymbol{\phi}_i^{\alpha_c},
\end{align}
for the case when $\sigma$ is just greater than $\sigma_c$, indicating that only the critical eigenvector $\boldsymbol{\phi^{\alpha_c}}$ significantly contributes to the perturbation and $u_i(t)$ will follow the shape of the critical mode  $\boldsymbol{\phi^{\alpha_c}}$. 
\begin{figure}[htbp]
\centering
\includegraphics[width=0.45\textwidth]{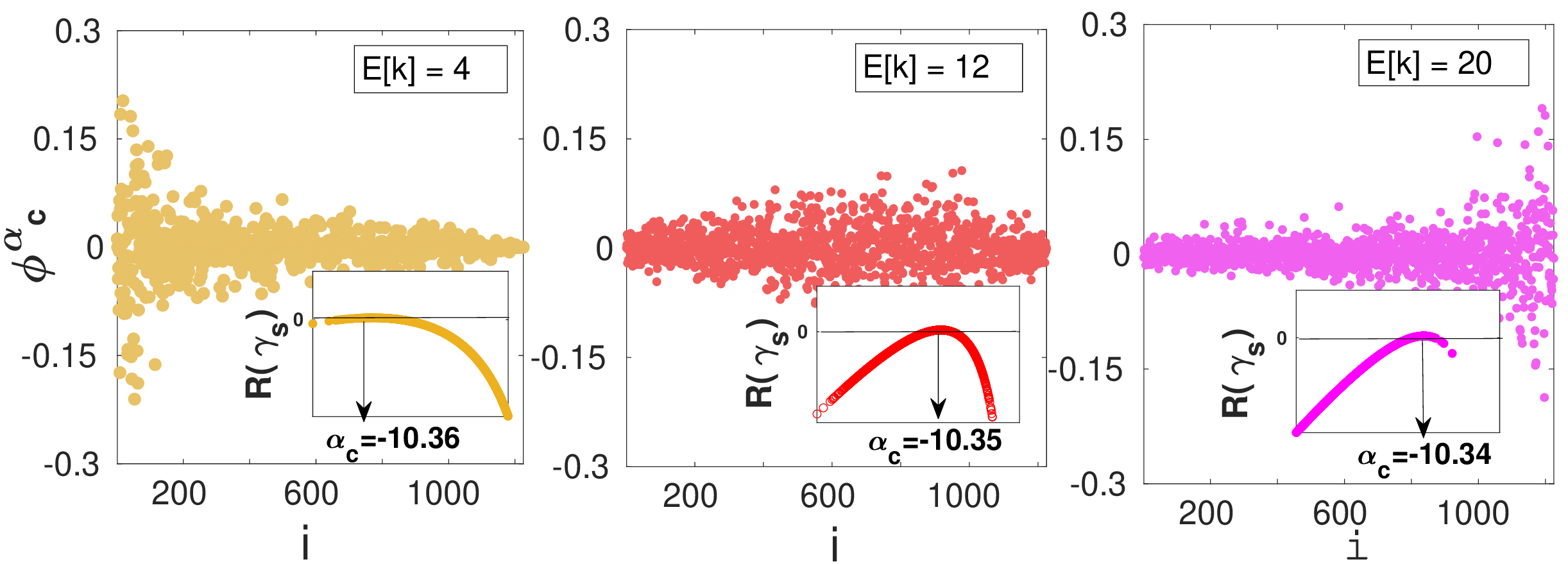}\\
\textbf{(a)} \hspace{0.8 in} \textbf{(b)} \hspace{0.8 in} \textbf{(c)}
\includegraphics[width=0.45\textwidth]{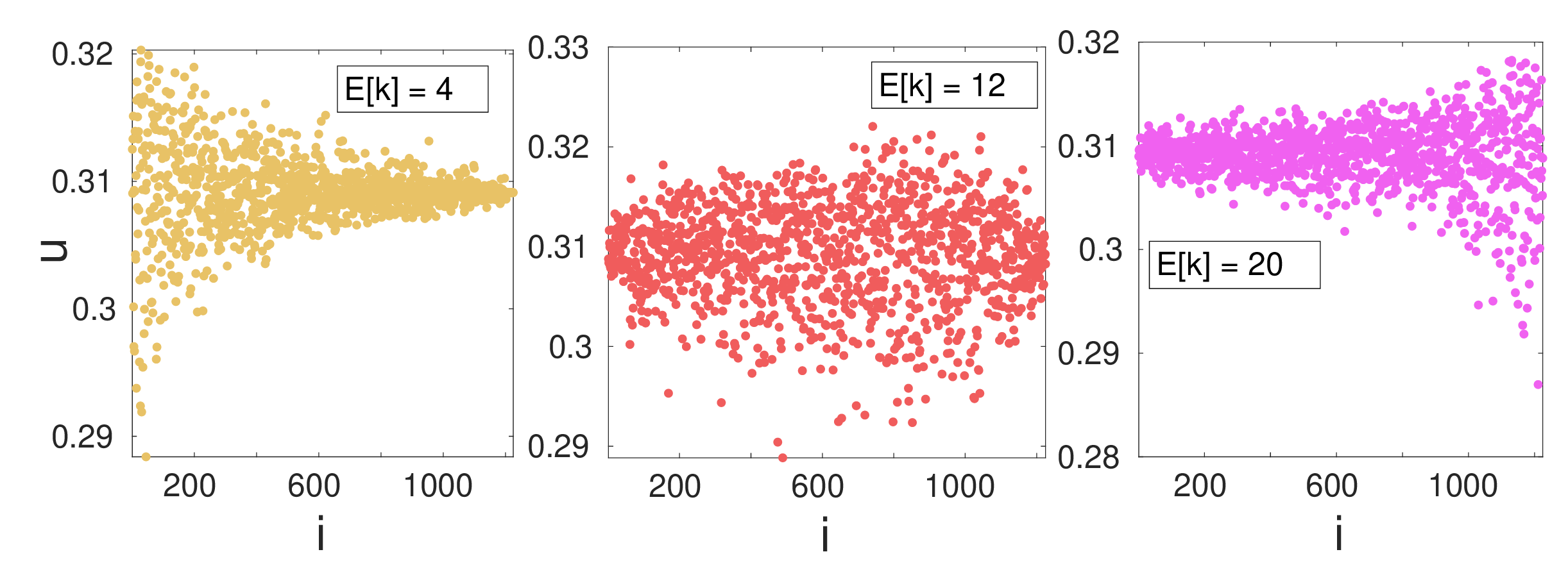}\\
\textbf{(d)} \hspace{0.8 in} \textbf{(e)} \hspace{0.8 in} \textbf{(f)}
    \caption{The critical case when $\sigma = 25.5$  (a)-(c): The critical Laplacian eigenvector $\boldsymbol{\phi^{\alpha_c}}$  as a function of nodes arranged in a descending order of degree for $E[k]$ equal to (a) $=4$, (b) $12$ and (c) $20$. Insets show the variation of $\Re(\gamma_s)$ as a function of $\alpha_s$. (d)- (f): Steady state values of $u_i$ as a function of nodes arranged in descending order of degree for three different cases of ER networks with $E[k]$ equal to (d)$4$, (e) $12$ and (f) $20$.}  
    \label{fig:eigenvector_critical}
\end{figure}
Figures \ref{fig:eigenvector_critical}(a), (b) and (c)  display the plot of $\boldsymbol{\phi^{\alpha_c}}$ for $E[k] = 4, 12, 20$ respectively, with $\sigma$ slightly above $\sigma_c$ at $25.5$. The nodes are arranged in descending order of their degrees the reasons for which will be explained later. The insets in these figures  show the variation of $\Re(\gamma_s)$ with $\alpha_s$, with $\alpha_c$ marked in each plot. 
Figure \ref{fig:eigenvector_critical}(d), (e) and (f) depict the steady-state values of $u$, with nodes arranged in descending order of their degrees, for $\sigma = 25.5$. Since $\sigma$ is slightly greater than $\sigma_c$, the patterns closely resemble the critical eigenvector $\boldsymbol{\phi}_i^{\alpha_c}$, as described in Eq.\eqref{perturb1}.}  

\subsubsection{Turing instability and the Laplacian eigenspectrum}

\begin{figure}[htbp]
\centering
\includegraphics[width=0.49\textwidth]{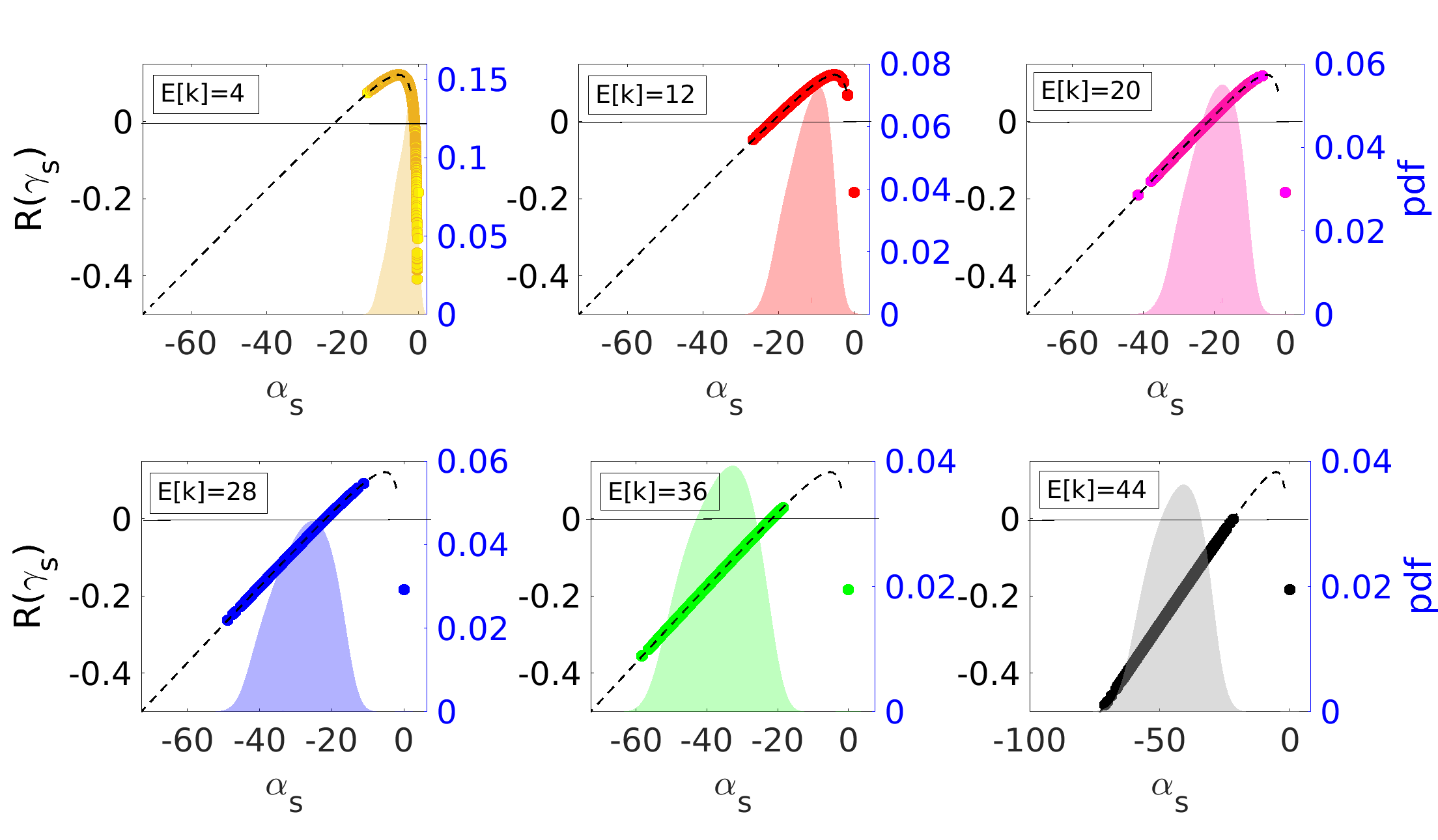}
    \caption{Variation of $\Re(\gamma_s)$ as a function of $\alpha_s$ for six different ER networks (left axis). Black dashed line represents Eq.\ref{gamma} for $\sigma= 120$. The colored dots represent the value of $\Re(\gamma_s)$ corresponding to specific values of $\alpha_s$.  Each sub-figure also shows the probability density function (pdf) of the eigenvalues of the corresponding  Laplacian matrix (right axis).} 
    \label{fig:eigendist2}
\end{figure}

\begin{figure}[htbp]
\centering
\includegraphics[width=0.49\textwidth]{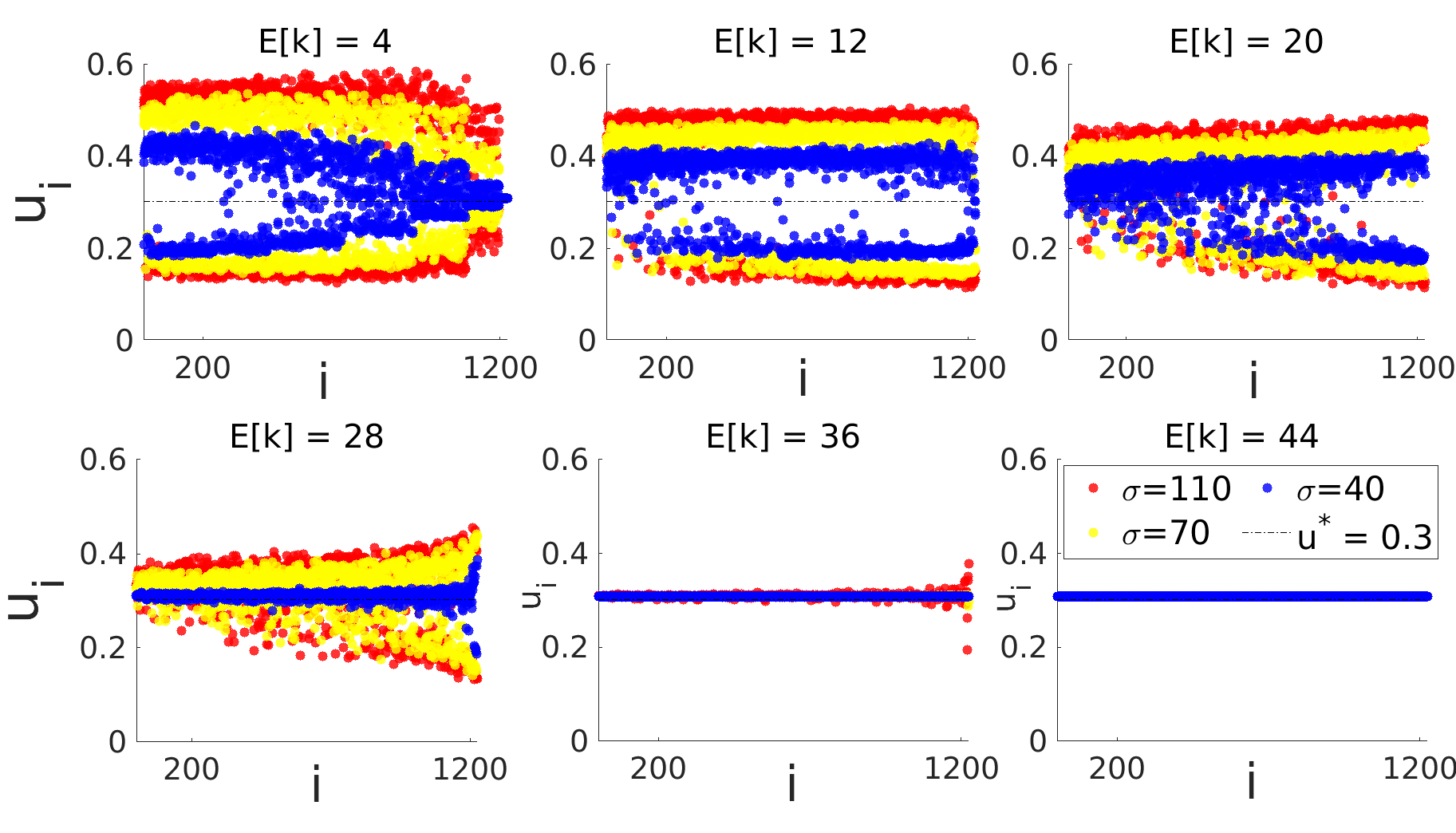}
    \caption{ER network with naturally low clustering: 
    Steady state values of $u$ for the different cases of ER networks considered with ${\rm E}[k]$ being the control parameter. The colours correspond to different cases of  $\sigma=40$ (blue), $\sigma=70$ (yellow) and $\sigma=110$ (red). The nodes are arranged in the descending order of their degree.}
    \label{fig:eigen2}
\end{figure}

While the discussions in Sec.\ref{background} indicates that $\sigma$ has to be greater than $\sigma_c$ for Turing instability, what is less clear that it  also depends on the eigenspectrum of the corresponding ${\bf L}$ of the network.
For the case $\sigma>\sigma_c$, as expected $\Re(\gamma_s)$ intersects the zero line at $\alpha_{s_1}$ and $\alpha_{s_2}$, with the intermediate region lying above zero. The dependence of the solution curve for Eq.\eqref{root} on the network topology is through $\alpha_s$ and hence Turing instability depends also on the eigenspectrum of ${\bf L}$. 
The close relationship between Turing instability and the Laplacian eigenvalue spectrum is illustrated through Fig. \ref{fig:eigendist2}, where the probability density function (pdf) of the eigenvalues and the $\Re(\gamma_s)$-$\alpha_s$ plots are shown for ER networks with varying ${\rm E}[k]$.
%
As expected, the eigenspectrum are shown to be centered about ${\rm E}[k]$. For the case ${\rm E}[k]=4$, the eigenspectrum is towards the right end and close to zero but as ${\rm E}[k]$ increases, the spectrum is observed to shift left with the eigenvalues becoming more negative. There exists one zero eigenvalue always (follows from the property of ${\bf L}$) and are shown as an isolated point in the spectrum. The range of the solution for $\Re(\gamma_s)$ for a network - denoted by the points lying on the solution curve of Eq.\eqref{root} - obviously coincides with the range of the spectrum. 
For the case ${\rm E}[k]=44$, Fig.\ref{fig:eigendist2} shows the eigenspectrum range to be such that all the values of $\Re(\gamma_s)<0$ indicating no instability and this is corroborated by the corresponding stationary values of $u_i$ in Fig.\ref{fig:eigen2}. 
Even for the case ${\rm E}[k]=36$, Fig.\ref{fig:eigendist2} shows only a few $\alpha_s$ lead to $\Re(\gamma_s)>0$ and this explains the weak instability  seen in the stationary $u_i$ values in Fig.\ref{fig:eigen2}. It follows  that $\sigma>\sigma_c$  is a necessary condition for Turing instability, but not a sufficient condition. For Turing instability, the range of the eigenspectrum of ${\bf L}$ should intersect with the range $[\alpha_{s_1},\alpha_{s_2}]$ also.

\subsubsection{Degree distribution and the Laplacian eigenspectrum}

It is interesting to note from Fig.\ref{fig:eigen2} that for the case of ${\rm E}[k]=4$, the stationary $u_i$ values show two branches towards the left end (higher degree nodes) and appear to merge together at the right end (lower degree nodes). For the case ${\rm E}[k]=12$ the two branches appear to be uniformly separated across all nodes. However, as ${\rm E}[k]$ is further increased, the two branches appear to merge for the higher degree nodes first, with the lower degree nodes merging only on further increasing of ${\rm E}[k]$. 

\begin{figure}[htbp]
\centering
\includegraphics[width=0.49\textwidth]{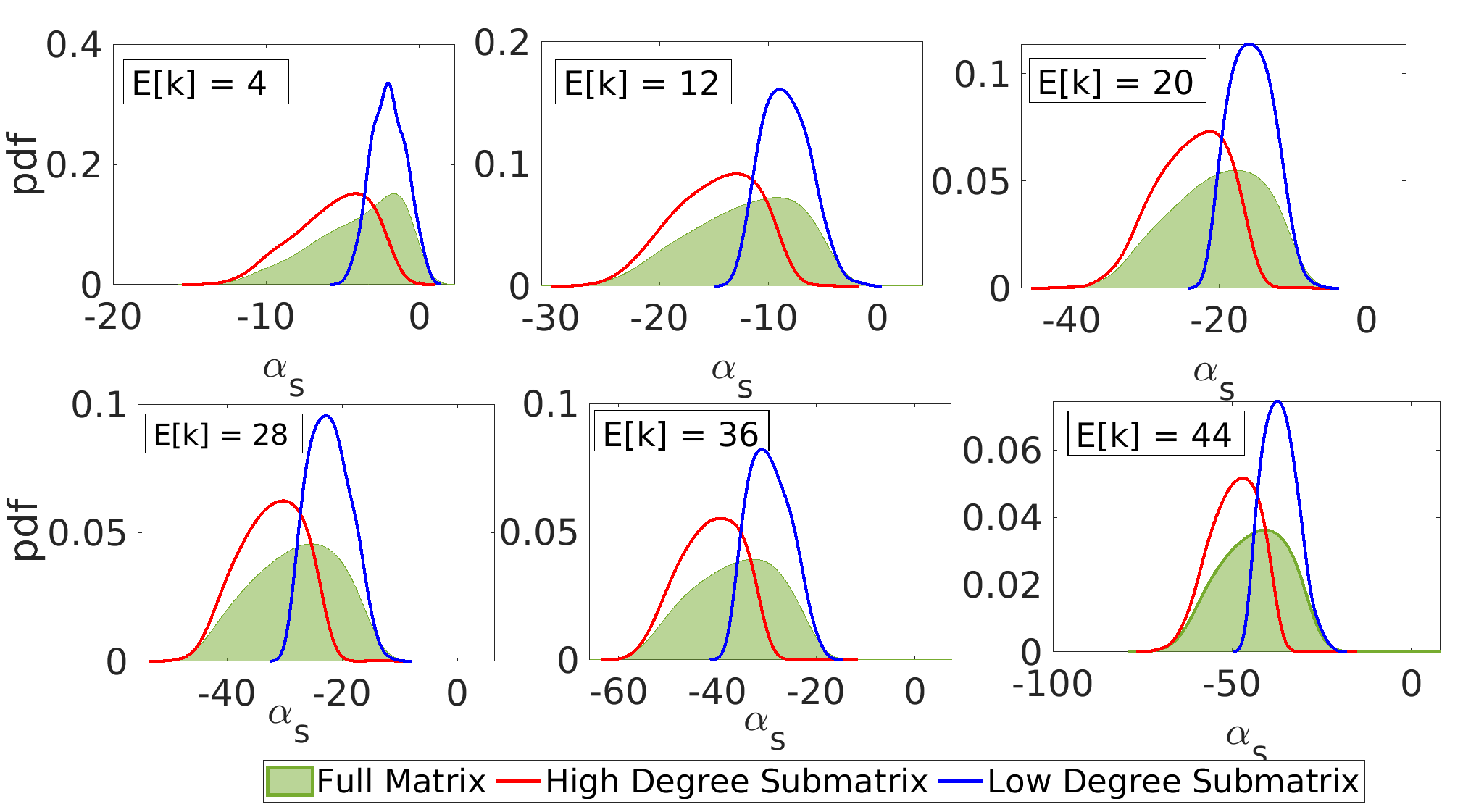}
    \caption{{Spectrum of the eigenvalues ($\alpha_s$) for the full laplacian matrix (in green), sub-matrix  of high degree nodes (red), and sub-matrix of low-degree nodes (blue). Full Laplacian matrix is of size $1225 \times 1225$, the high-degree sub-matrix is  $800 \times 800$ and the low-degree sub-matrix is $425 \times 425$.}}
    \label{fig:eigen_high_low}
\end{figure}

{Correlating the observations with the eigenspectra shown in Fig.\ref{fig:eigendist2} and noting that the stationary $u_i$ values are obtained from contributions of the unstable eigenmodes (see Eq.\eqref{pert1}), it seems to indicate that the higher degree nodes contribute towards the left end eigenvalues of the spectrum and the lower degree nodes are associated with the right end of the spectrum. 
To test this hypothesis, the Laplacian matrix \textbf{L} is rearranged in descending order of node degrees. Interchanging the rows and columns of a matrix constitutes a similarity transformation, which does not alter the eigenvalues of the rearranged matrix. Next, two submatrices are constructed from the original matrix. The first submatrix is formed by selecting the first 800 rows and columns of the rearranged matrix, referred to as the high degree submatrix. The remaining rows and columns form the low degree submatrix. The number 800 is chosen based on observations from Fig.\ref{fig:eigen2} for $E[k]=4$, as the two branches are observed to merge together at approximately node number 800. 
Figure \ref{fig:eigen_high_low} shows the eigenspectra of $\alpha_s$ for ${\bf L}$, high-degree submatrix, and low-degree submatrix represented by green, red and blue curves respectively, for various cases of ER network with different $E[k]$.}
{It is observed that the eigenspectrum comprising of the higher degree nodes are always shifted towards the left of the original eigenspectrum, while the corresponding eigenspectrum for the lower degree nodes gets shifted towards the right. This observation, found to be consistent across all the different network topologies shown in Fig.\ref{fig:eigen_high_low} seems to lend credence that indeed the higher degree nodes contribute towards the left end of the eigenspectrum and vice versa.}  

This observation can be further investigated with the  
help of eigenvalue perturbation theory\cite{rellich1969perturbation,li2006matrix,hogben2006handbook}. 
Consider the eigenvalue problem, 
\begin{align}
\label{tag1}
\mathbf{L} \boldsymbol{\phi}_i = \alpha_i  \boldsymbol{\phi}_i
\end{align}
where \(\mathbf{L}\) is a matrix with significantly larger diagonal elements compared to its off-diagonal elements, thus making \(\mathbf{L}\) a diagonally dominant matrix. According to matrix perturbation theory, \(\mathbf{L}\) can be decomposed into the sum of a diagonal matrix \(\mathbf{D}\) and a matrix \(\mathbf{A}\) with relatively small entries as
\begin{align}
\label{tag2}
\mathbf{L} = \mathbf{D} + \mathbf{A}.
\end{align}
In the context of the problem being discussed, \(\mathbf{L}\) is the Laplacian matrix, \(\mathbf{D}\) is a diagonal matrix with the degree \(-k_i\) as its diagonal elements, and \(\mathbf{A}\) being the adjacency matrix. It is further assumed that the  eigenvalue problem for ${\bf D}$ can be written as 
\begin{equation}
    \bf{D} \boldsymbol{\phi_0} = \alpha_0  \boldsymbol{\phi_0}. 
\end{equation}
For a network with sufficiently high value for ${\rm E}[k]$, the elements of \(\mathbf{A}\) are clearly significantly  smaller than \(\mathbf{D}\) and hence, the new eigenvalues $\alpha_i$ and eigenvectors $\phi_i$ corresponding to ${\bf L}$ can be assumed to be obtained in terms of small perturbation with respect to $\bf{D}$. Mathematically, this implies
\begin{align}
\alpha_i = \alpha_{0i} + \delta \alpha_i \quad \text{and} \quad \boldsymbol{\phi}_i = \boldsymbol{\phi}_{0i} + \delta \boldsymbol{\phi}_i,
\label{tag00}
\end{align}
where $\delta \alpha_i$ and $\delta {\boldsymbol \phi}_i$ are the perturbations for the eigenvalue and the eigenvector. Substituting Eq.\eqref{tag2} on the left side of Eq.\eqref{tag1} and Eq.\eqref{tag00} on its right side,  
the eigenvalue problem can be written as
\begin{align}
(\mathbf{D} + \mathbf{A})(\boldsymbol{\phi}_{0i} + \delta \boldsymbol{\phi}_i) = (\alpha_{0i} + \delta \alpha_i)(\boldsymbol{\phi}_{0i} + \delta \boldsymbol{\phi}_i).
\end{align}
Expanding both sides and ignoring higher-order terms (\(\delta \boldsymbol{\phi}_i \delta \alpha_i\)) leads to
\begin{align}
& \mathbf{D} \boldsymbol{\phi}_{0i} + \mathbf{D} \delta \boldsymbol{\phi}_i + \mathbf{A} \boldsymbol{\phi}_{0i} + \mathbf{A} \delta \boldsymbol{\phi}_i \nonumber \\
& = \alpha_{0i} \boldsymbol{\phi}_{0i} + \alpha_{0i} \delta \boldsymbol{\phi}_i + \delta \alpha_i \boldsymbol{\phi}_{0i} + \delta \alpha_i \delta \boldsymbol{\phi}_i.
\end{align}

This equation can be further simplified by cancelling equivalent terms from both sides,  leading to the following simplified form
\begin{align}
(\mathbf{D} - \alpha_{0i} \mathbf{I}) \delta \boldsymbol{\phi}_i = \delta \alpha_i \boldsymbol{\phi}_{0i} - \mathbf{A} \boldsymbol{\phi}_{0i}.
\label{tag000}
\end{align}
Multiplying both sides by \(\boldsymbol{\phi}_{0i}^T\), 
and since \(\boldsymbol{\phi}_{0i}^T (\mathbf{D} - \alpha_{0i} \mathbf{I}) = 0\), Eq.\eqref{tag000} simplifies to 
\begin{align}
0 = \delta \alpha_i \boldsymbol{\phi}_{0i}^T \boldsymbol{\phi}_{0i} - \boldsymbol{\phi}_{0i}^T \mathbf{A} \boldsymbol{\phi}_{0i}.
\end{align}
Thus, the first-order perturbation correction to the eigenvalues is given by
\begin{align}
\delta \alpha_i = \boldsymbol{\phi}_{0i}^T \mathbf{A} \boldsymbol{\phi}_{0i}
\end{align}
The upper bound for $\delta \alpha_i$ can be estimated using the  Bauer–Fike theorem\cite{bauer1960norms}.\\
\begin{theorem}[Bauer-Fike Theorem]
{\it Let \( \mathbf{D} \) be a diagonalizable \( N \times N \) matrix, and  \( \mathbf{L} = \mathbf{D} + \mathbf{A} \) where \( \mathbf{A} \) is any \( N \times N \) matrix. Also, assume that \( \mathbf{D} = \mathbf{V} \mathbf{\xi} \mathbf{V}^{-1} \) be the diagonalization of  \( \mathbf{D} \), where \( \mathbf{V} \) is the matrix of eigenvectors of \( \mathbf{D} \) and \( \mathbf{\xi} \) is the diagonal matrix of eigenvalues of \( \mathbf{D} \). Then, for any eigenvalue \( \alpha \) of \( \mathbf{L} \), there exists an eigenvalue \( \alpha_0 \) of \( \mathbf{D} \) such that:
\begin{align}
|\alpha- \alpha_0| \leq \chi_p(\mathbf{V}) \| \mathbf{A} \|_p,
\end{align}
where \( \chi_p(\mathbf{V}) = \|\mathbf{V}\|_p \|\mathbf{V}^{-1}\|_p \) is the condition number in p-norm of the matrix of eigenvectors \( \mathbf{V} \) of \( \mathbf{D} \). }\\
\end{theorem}
For the diagonal matrix $\bf{D}$, the eigenvector matrix is the identity matrix, which implies 
$ \chi_p(\mathbf{V}) =  \|\mathbf{V}\| \|\mathbf{V}^{-1}\| = \|\mathbf{V}\| \|\mathbf{V}^{T}\| = 1$. Since \( \chi_p(\mathbf{V}) = 1 \), the perturbation to the eigenvalue \( \alpha_i \) is bounded by the norm of \( \mathbf{A} \) i.e., $| \delta \alpha_i | \leq \| \mathbf{A} \|_p$. Essentially, the Bauer-Fike theorem tells that the eigenvalues of a perturbed matrix \(\mathbf{L}\) remain stable and closely approximate those of the original diagonal matrix \(\mathbf{D}\), with deviations from these original eigenvalues constrained by the magnitude of the perturbation matrix \(\mathbf{A}\). Since \(\mathbf{D}\) is a diagonal matrix with its diagonal elements representing the degrees \(-k_i\) of the network, and for diagonal matrices, the eigenvalues coincide with the diagonal elements\cite{strang2012linear}, the eigenvalues of the Laplacian matrix \(\mathbf{L}\) are approximately equal to the degrees of the network, i.e., \(\alpha_i \approx -k_i\).  

\begin{figure*}[t]
\centering
\includegraphics[width = 0.33\textwidth]{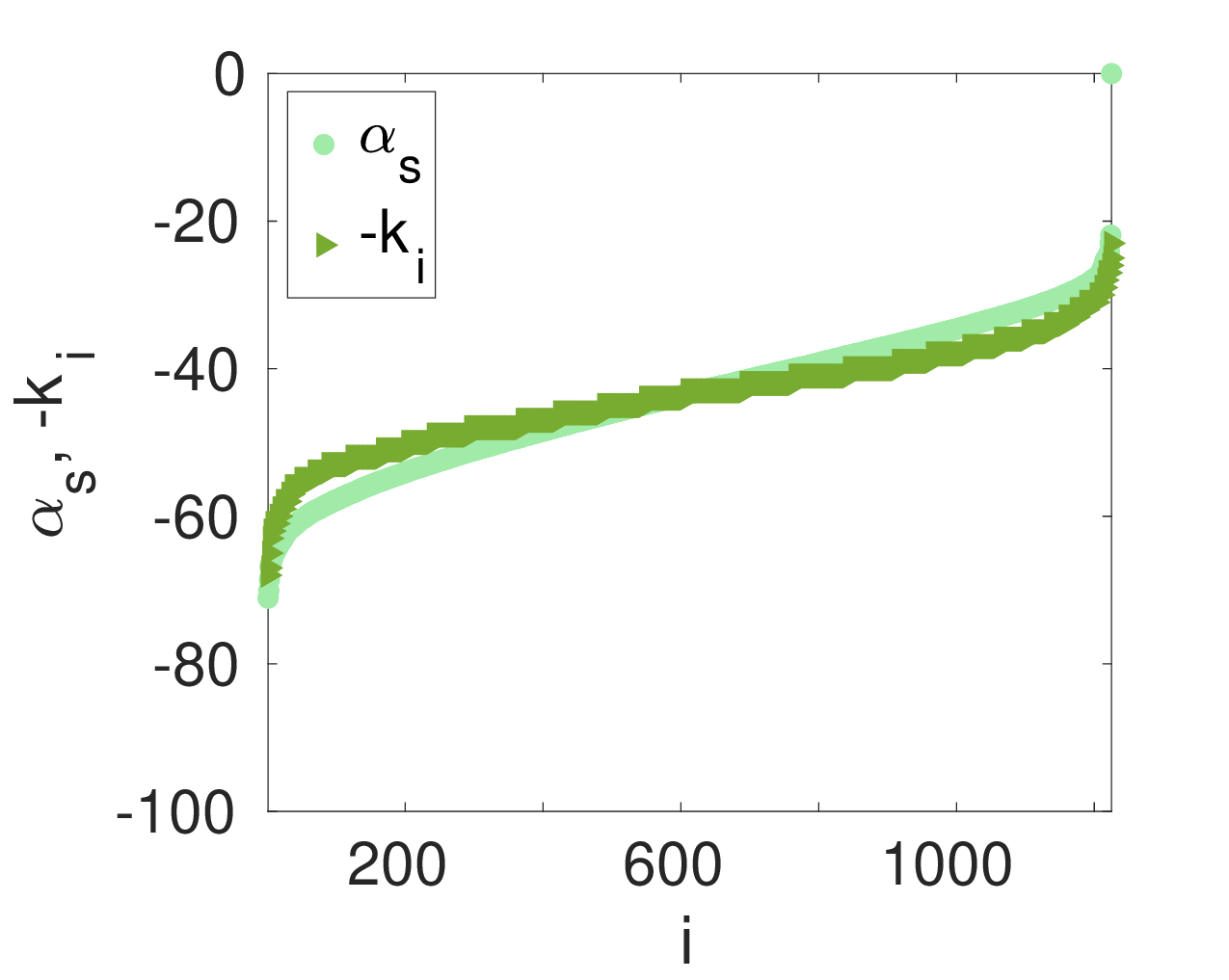}
\includegraphics[width = 0.33\textwidth]{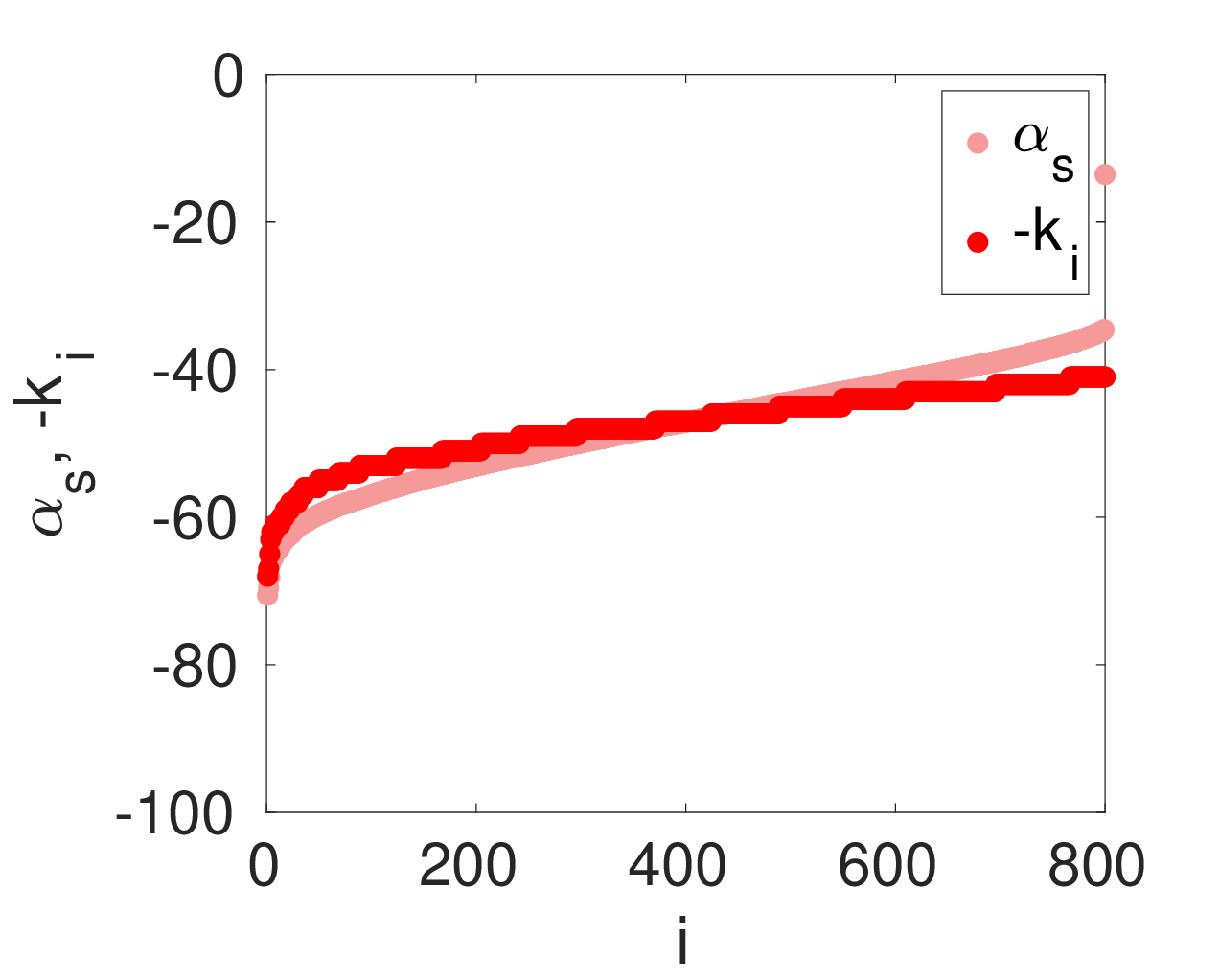}
\includegraphics[width = 0.33\textwidth]{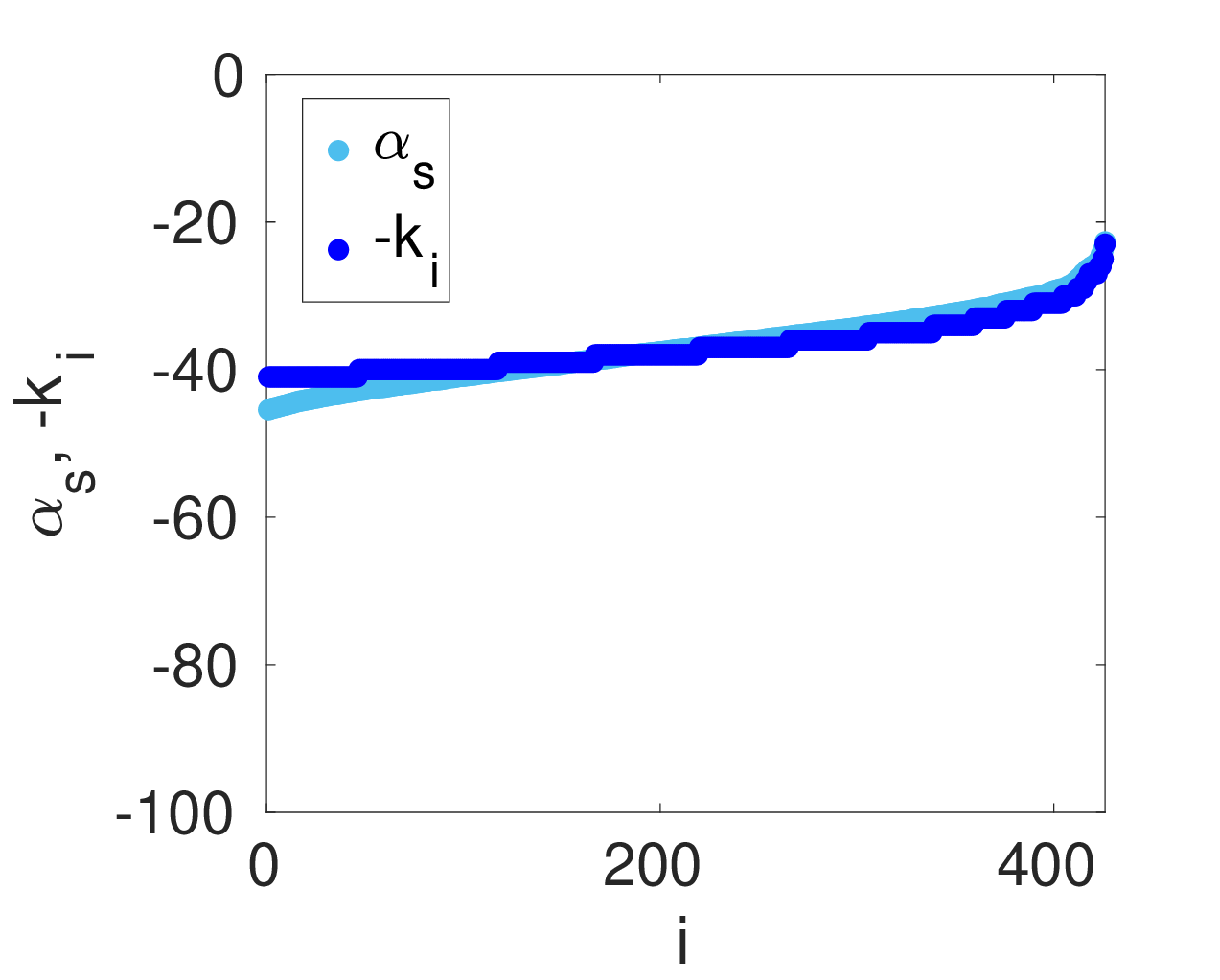}
    \textbf{(a)} \hspace{1.9 in} \textbf{(b)} \hspace{1.9 in} \textbf{(c)}
    \caption{{Variation of $\alpha_s$ and $-k_i$ for $E[k]=44$, with nodes arranged in descending order of degree,  for (a) full network ($1225 \times 1225$) (b) network with the first 800 higher degree nodes 
    and (c) network with the $425$ lowest degree nodes. 
    The eigenvalues of $\bf{L}$ are approximately equal to $-k_i$ in all three cases.}}
    \label{fig:comparison_k_alpha}
\end{figure*}

This result is  validated through numerical simulations, the results of which are are shown in Fig.\ref{fig:comparison_k_alpha}, where the variation of  $-k_i$ (negative nodal degrees) as well as the eigenvalues of ${\bf L}$ are shown in terms of the nodes, when arranged in descending order of degree. 
Figure \ref{fig:comparison_k_alpha}(a) shows this variation for the ER network with ${\rm E}[k]=44$.  It is observed that the two plots follow almost identical trend, with the deviations being within $5\%$. This indicates that the degree distribution and the eigenspectrum are almost identical.  
%
%
Note that ${\bf L}$ was constructed after a similarity transformation by which the rows and columns were arranged in descending orders of nodal degree. Two submatrices were subsequently constructed from ${\bf L}$, with ${\bf L}_1$ being constructed from the first 800 rows and columns of ${\bf L}$ and ${\bf L}_2$ being constructed from the remaining 425 rows and columns. Thus, ${\bf L}_1$ consists of nodes with higher degrees while ${\bf L}_2$ consist of nodes with lower degrees. Similar plots as Fig.\ref{fig:comparison_k_alpha}(a) were constructed for both ${\bf L}_2$ and ${\bf L}_3$; see Figs.\ref{fig:comparison_k_alpha}(b)-(c). In these cases as well, one can see that the eigenvalues $\alpha_s$ and the eigenspectrum follow identical trends, indicating that the nodal degree and the eigenvalues of the Laplacian are almost equal and have a one-to-one mapping. This also explains the patterns of the variation of the steady state values of $u_i$ in Fig.\ref{fig:eigen2}. This aspect does not appear to have been discussed in the literature earlier.

\subsection{ER network with high clustering}
In this section, the effect of high clustering on Turing instability in ER networks is investigated. The algorithm ``ClustRNet"\cite{bansal2009exploring}  allows for the generation of networks with a specified clustering coefficient, without changing the degree distribution (implying no change in ${\rm E}[k]$ either). 
The underlying principle of this algorithm is that it  rewires edges within the network iteratively, to introduce triangles among the nodes. This process, known as edge rewiring or edge swapping, is a widely used method for generating networks with desired characteristics \cite{gale1956theorem, maslov2002specificity}. Each rewiring takes a set of five connected nodes   and swaps the outer edges.
\begin{figure}[htbp]
\centering
\includegraphics[width=0.49\textwidth]{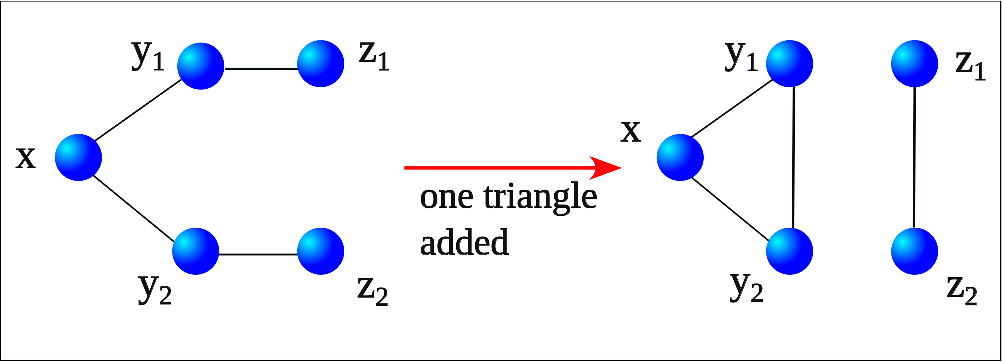}\\
\textbf{(a)}\\
\includegraphics[width=0.49\textwidth]{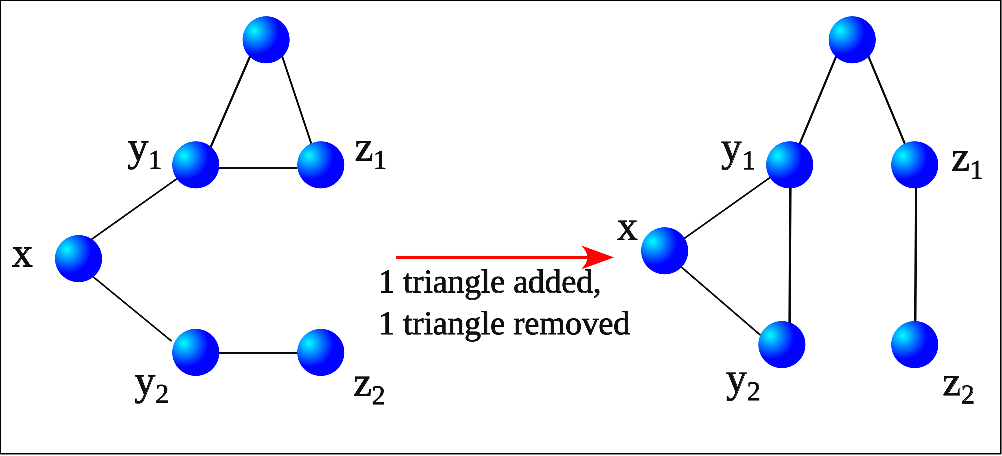}\\
 \textbf{(b)}
    \caption{(a) The configuration of edges before (left) and after (right) a rewiring step. The rewiring introduces a triangle among nodes {$(x, y_1, y_2)$} (b) As the rewiring does not alter the number of triangles, this is rejected by the algorithm }
    \label{fig:rewiring}
\end{figure}
 For example, in Fig. \ref{fig:rewiring}(a), the 5 nodes $(x, y_1, y_2, z_1, z_2)$ are connected through the 
four edges {$(x,y_1), (x, y_2), (y_1 , z_1), (y_2 , z_2)$}. Through edge swapping,  the outer edges $(y_1,z_1)$ and $(y_2, z_2)$ are swapped to form new edges $(z_1, z_2)$ and $(y_1,y_2)$. This introduces a triangle among nodes $(x, y_1, y_2)$, without affecting the degree of any of the nodes. If the rewiring does not alter the number of triangles, as shown in Fig.\ref{fig:rewiring}(b), the scenario is rejected by the algorithm. Rewiring of edges is carried out until the desired level of clustering is achieved.

\begin{figure}[htbp]
\centering
\includegraphics[width=0.49\textwidth]{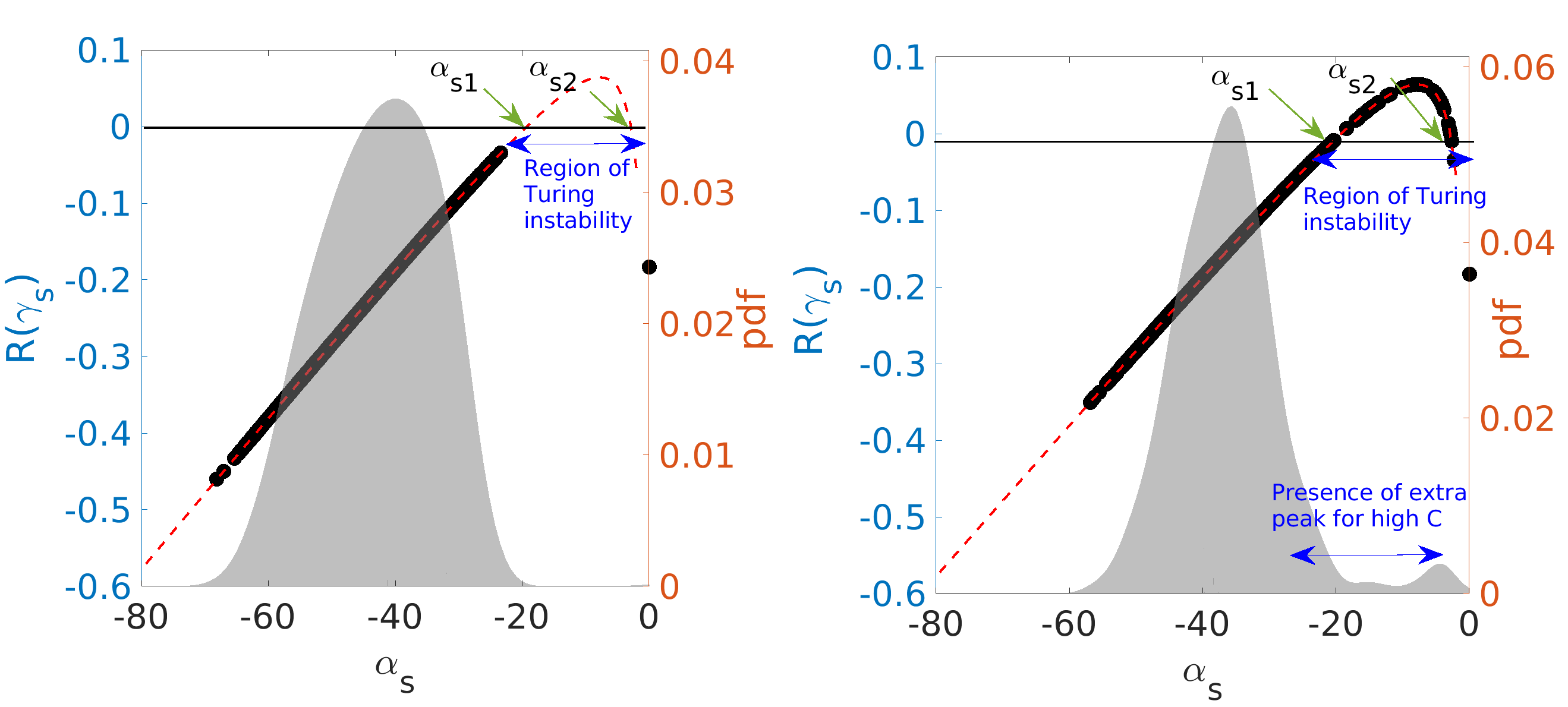}\\
\textbf{(a)} \hspace{1.8 in} \textbf{(b)}
    \caption{{Variation of $\Re(\gamma_s)$ as a function of $\alpha_s$; ER network with $E[k]=44$, $\sigma=70$; (a) $C=0.034$ and (b) $C=0.6$. The eigenspectrum for the corresponding  ${\bf L}$ are shown in grey.}}
    \label{fig:dispersion_curve_ER}
\end{figure}

The effect of clustering is next investigated for an ER network with ${\rm E}[k]=44$. The naturally occurring network (case 1) has $C=0.034$, which indicates low clustering. Using the edge swapping algorithm, a rewired ER network is constructed with $C=0.6$ (case 2), indicating a clustering that is almost 20 times higher. The eigenvalue spectrum of ${\bf L}$ shown in Fig.\ref{fig:diff_cc_spectra_ER} corresponds to the rewired networks. Here, Fig.\ref{fig:diff_cc_spectra_ER}(a) being the case of the naturally occurring network, while Fig.\ref{fig:diff_cc_spectra_ER}(d) corresponds to the rewired network with the desired C. 
Figure \ref{fig:dispersion_curve_ER}(a) and (b) shows the variation of $\Re(\gamma_s)$ as a function of $\alpha_s$ for these two cases. 
{In the former case, even though $\Re(\gamma_s)>0$ for the selected value of $\sigma$, the network topology is such that the Laplacian eigenspectrum have no intersection with the interval $[\alpha_{s_1},\alpha_{s_2}]$ and hence no Turing instability is observed; see Fig.\ref{fig:comparison_ER}(a) where it can be seen that all the nodes attain the steady state value $u^*$.
\begin{figure}[htbp]
\centering
\includegraphics[width=0.48\textwidth]{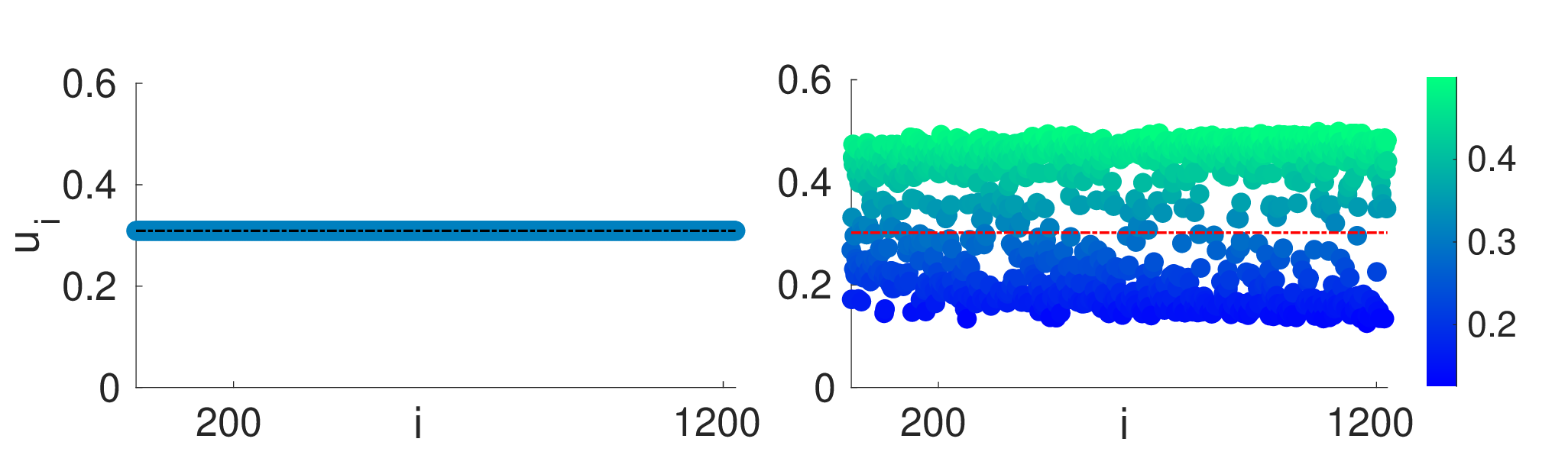}\\
\textbf{(a)} \hspace{2.0 in} \textbf{(b)}\\
\includegraphics[width=0.48\textwidth]{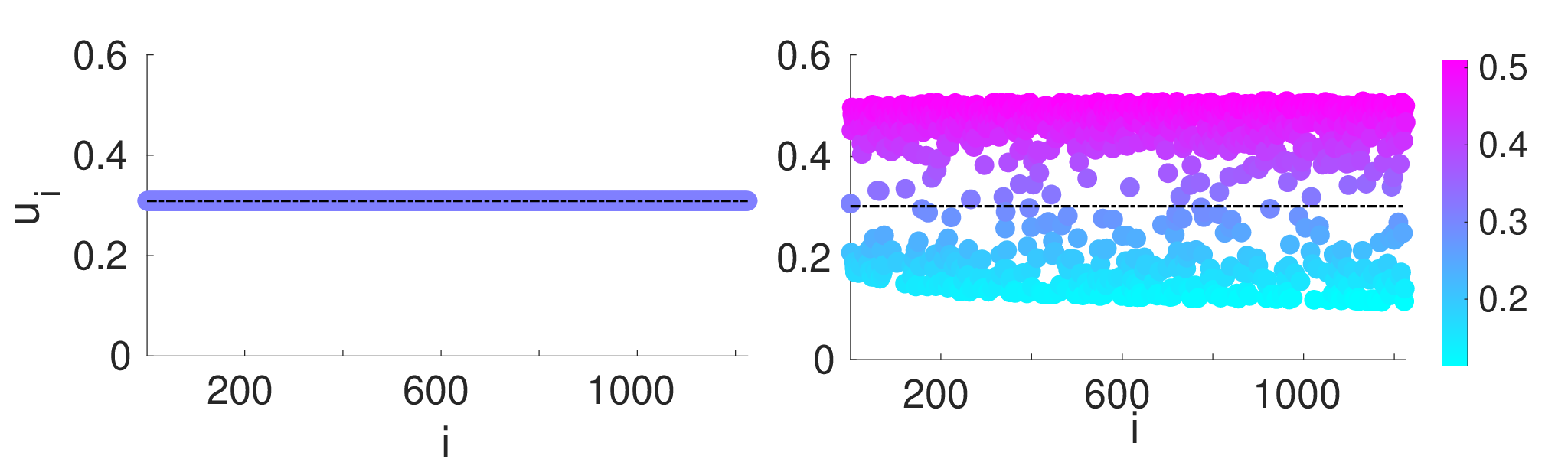}\\
\textbf{(c)} \hspace{2.0 in} \textbf{(d)}
    \caption{Steady state values of $u$ plotted for $E[k]=44$ (a) $C=0.034$, $\sigma=70$ (b) $C=0.6$, $\sigma=70$ (c) $C=0.034$, $\sigma=110$, (d) $C=0.6$, $\sigma=110$}
    \label{fig:comparison_ER}
\end{figure}
On the other hand, Fig.\ref{fig:dispersion_curve_ER}(b) shows the same results for an ER network with identical degree distribution and ${\rm E}[k]$, but suitably rewired such that its $C$ is now 20 times higher. The eigenspectrum is now observed to have multiple peaks towards zero. As a result, some of the Laplacian eigenvalues $\alpha_s$ lie within the interval $[\alpha_{s_1},\alpha_{s_2}]$, resulting in Turing instability as can be seen from Fig.\ref{fig:comparison_ER}(b). The significant effect of clustering is evident from Figs. \ref{fig:comparison_ER}(c)-(d) as well, where even though $\sigma$ is significantly high, no instability is observed in Fig.\ref{fig:comparison_ER}(c) where $C=0.034$ but observed in Fig.\ref{fig:comparison_ER}(d), where $C=0.6$. These results highlight the importance of clustering for Turing instability.

\subsection{Scale free network}
The effects of clustering on Turing instability in a scale free network is investigated next. A scale-free network is simnulated following the Barabasi-Albert algorithm.
%
Subsequently, the network Laplacian of a given BA network is rewired until the desired level of clustering is achieved using the ``ClustRNet" algorithm. 
\begin{figure}[htbp]
\centering
\includegraphics[width=0.49\textwidth]{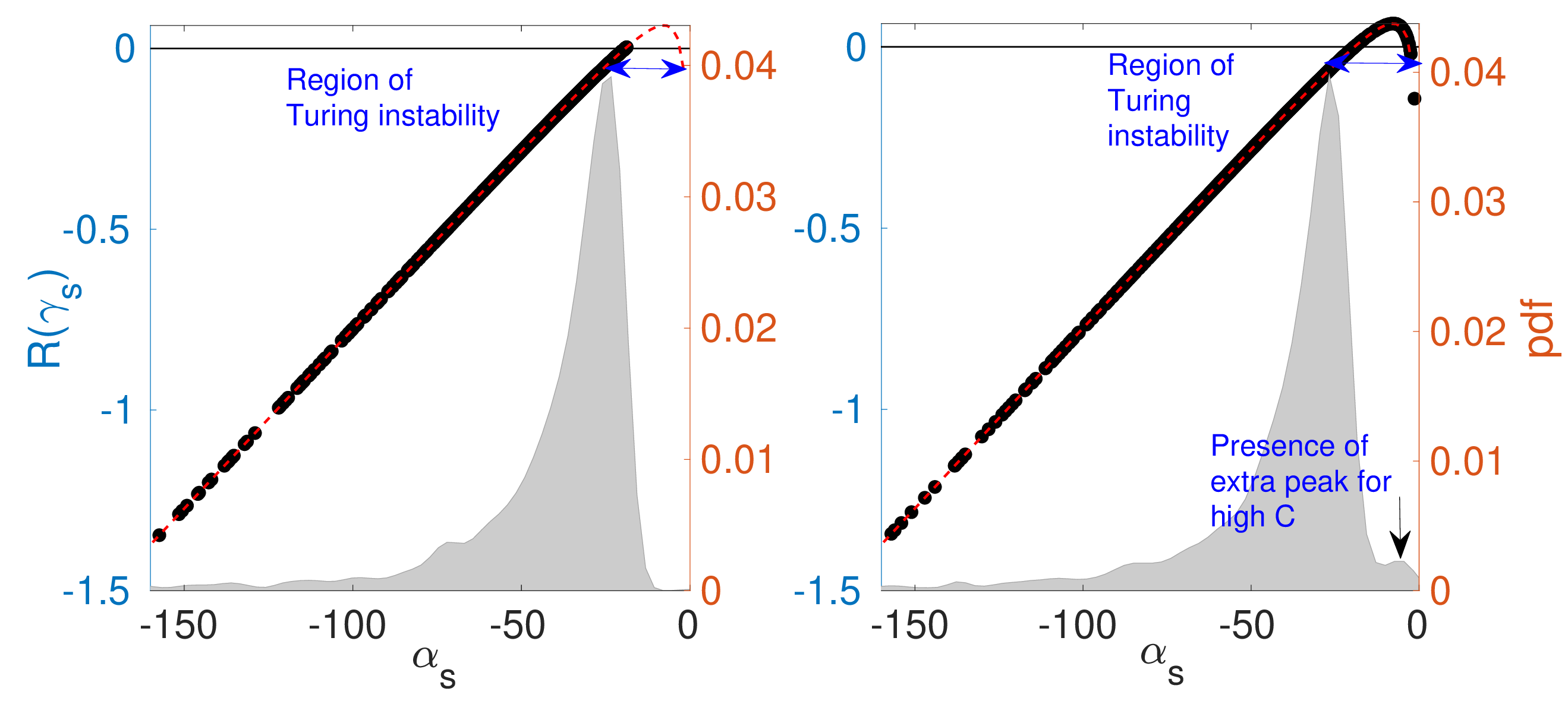}\\
\textbf{(a)} \hspace{2.4 in} \textbf{(b)}
    \caption{Variation of $\Re(\gamma_s)$ with $\alpha_s$ and the corresponding eigenspectrum for scale free network with  $E[k]=44$: (a) networks simulated using BA algorithm leading to $C=0.095$, (b) rewired scale free network with  $C=0.6$.} 
    %
    \label{fig:dispersion_curve_BA}
\end{figure}
The original BA network with $E[k] = 44$  exhibits low clustering with a clustering coefficient \( C = 0.095 \). By employing the edge-swapping algorithm, the clustering coefficient is increased to $C = 0.6$ . Figures \ref{fig:dispersion_curve_BA}(a) and (b) respectively show the eigenspectra of $\alpha_s$  for the original  and the rewired networks. As in the case of ER networks, the rewired scale-free network with a higher clustering shows an additional peak towards zero. The spectra in both these cases show a long left tail which is reflective of the power-law degree distribution. 


Both these figures also show the plot of $\Re(\gamma_s)$ as a function of $\alpha_s$, as obtained from the solution of Eq.\eqref{root}; see the dashed line. The circles on these lines represent the actual values of $\Re(\gamma_s)$ obtained for the $\alpha_s$ for the corresponding Laplacian matrix, and hence takes into account the topology of the network. It cane be seen In Fig.\ref{fig:dispersion_curve_BA}(a) that all the eigenmodes lie below zero and hence does not lead to Turing instability. On the other hand, the $\Re(\gamma_s)$ values for the rewrired network has several points in the $\Re(\gamma_s)-\alpha_s$ plot that lie above zero; see Fig.\ref{fig:dispersion_curve_BA}(b). These are the consequence of the eigenvalues of the Laplacian of the rewired scale-free network that are now close to zero, indicating the presence of several unstable eigenmodes that contribute to Turing instability. Thus, the observations that were made with respect to the ER networks earlier are also found to be valid for  scale-free networks.

\subsection{$\mathbb{S}^1$ / $\mathbb{H}^2$ network}
To illustrate the role played by clustering on Turing patterns, in this section Holling type III - Leslie Gower dynamics is investigated in geometric soft configuration model. A recent study \cite{van2023emergence} on Turing instability  on $\mathbb{S}^1$ / $\mathbb{H}^2$ networks has demonstrated that the eigenvectors corresponding to eigenvalues close to zero exhibit a well-defined periodic structure in the similarity space. In the realm of geometric networks, a Turing pattern is characterized by the distinct periodic arrangement of the pattern. Since it is established that $\delta u_i$ and $\delta v_i$ are proportional to $\bf{\phi_i^{s_c}}$, it is crucial to select parameters $D$ and $\sigma$ in a manner that ensures the eigenvector displaying periodic structure reaches a critical state. 
Since $\sigma_c = 25.049$, the value of $\sigma$ is set slightly above $\sigma_c$ at $25.1$. Unstable eigenvalues consequently fall within the interval $[\alpha_{s1},\alpha_{s2}]$, where $\alpha_{s1} = 0.0741/D$ and $\alpha_{s2} = 0.1043/D$ as determined from Eq.\eqref{eigen}. The most unstable eigenvalue is determined as $\alpha_{max}=0.1364/D$ from Eq.\eqref{eigen_max}. The eigenvalue $\alpha^*$, corresponding to the eigenvector displaying periodic structure, is set to $\alpha_{max} = \alpha^*$, and the value of $D$ is calculated. In the current study, $\alpha_3$ is taken to be the eigenvalue of interest $\alpha^*$.
\begin{figure}[htbp]
\centering{
\includegraphics[width=0.4\textwidth]{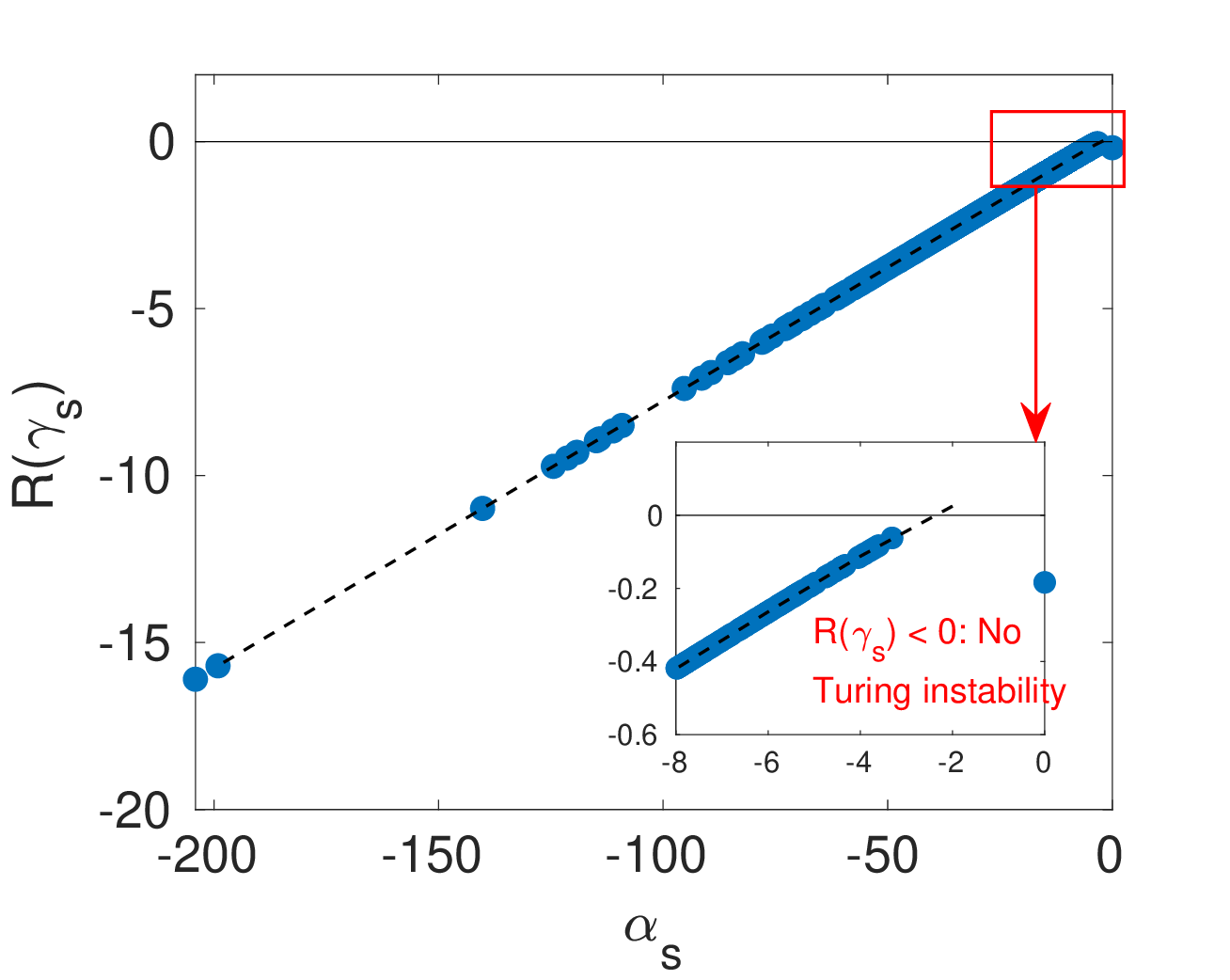}\\
\textbf{(a)}\\
\includegraphics[width=0.4\textwidth]{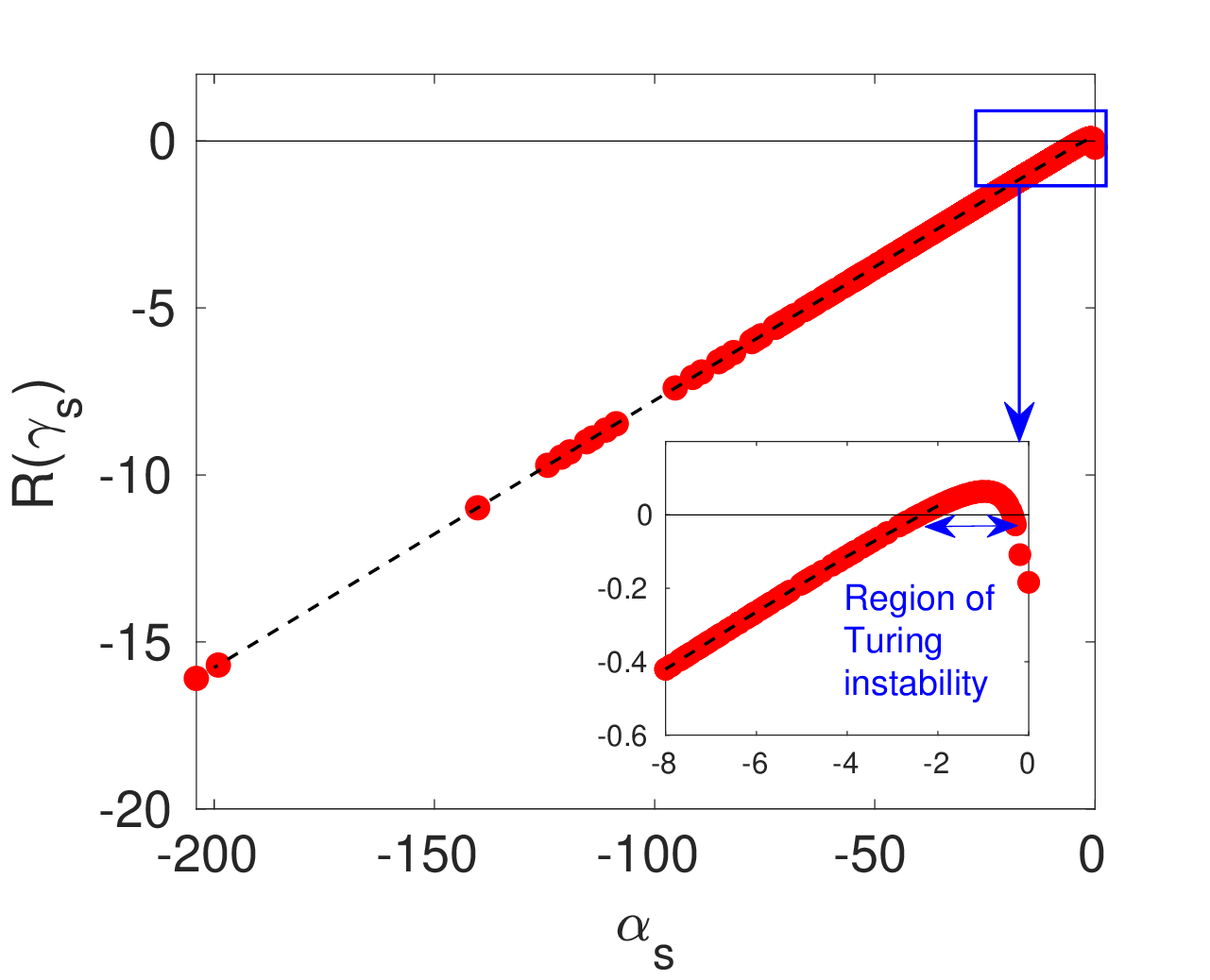}\\
\textbf{(b)}}
    \caption{Variation of $\Re(\gamma_s)$ with $\alpha_s$ in $\mathbb{S}^1$ / $\mathbb{H}^2$ network with parameters $(N,E[k], \beta, \Lambda)$: (a) (1225, 50, 1.1, 2.86)  (b) (1225, 50, 2.51, 2.86). The inset provides a closer view of the region around zero for $\alpha_s$.}
    \label{fig:hyperbolic_eigendist}
\end{figure}
Figs. \ref{fig:hyperbolic_eigendist}(a) and (b) illustrate the variation of $\Re(\gamma_s)$ with $\alpha_s$ in $\mathbb{S}^1/\mathbb{H}^2$ networks, with the parameters $(N, E[k], \beta, \Lambda) = (1225, 50, 1.1, 2.86)$ and $(1225, 50, 2.51, 2.86)$ respectively. 
The parameter $\beta$ controls the clustering in the network.  $\beta = 1.1$ results in low clustering with $C= 0.0935$ and  $\Re(\gamma_s) < 0$ for all $\alpha_s$ corresponding to the Laplacian eigenspectrum (see Fig.\ref{fig:hyperbolic_eigendist}(a)) indicating no Turing instability. 
A highly clustered network is obtained for $\beta=2.51$ with $C=0.573$. Figure \ref{fig:hyperbolic_eigendist}(b) shows that $\Re(\gamma_s) > 0$ for $\alpha_s$ close to 0 (see inset) 
indicating Turing instability. These observations are consistent with those in ER and scale free networks. 

The underlying geometric network topology for these two networks is shown in Figs.\ref{fig:hyperbolic_pattern}(a) and (e) respectively, where the nodes are located according to their coordinate in the hyperbolic space. 
\begin{figure}[htbp]
\centering
\includegraphics[width=0.45\textwidth]{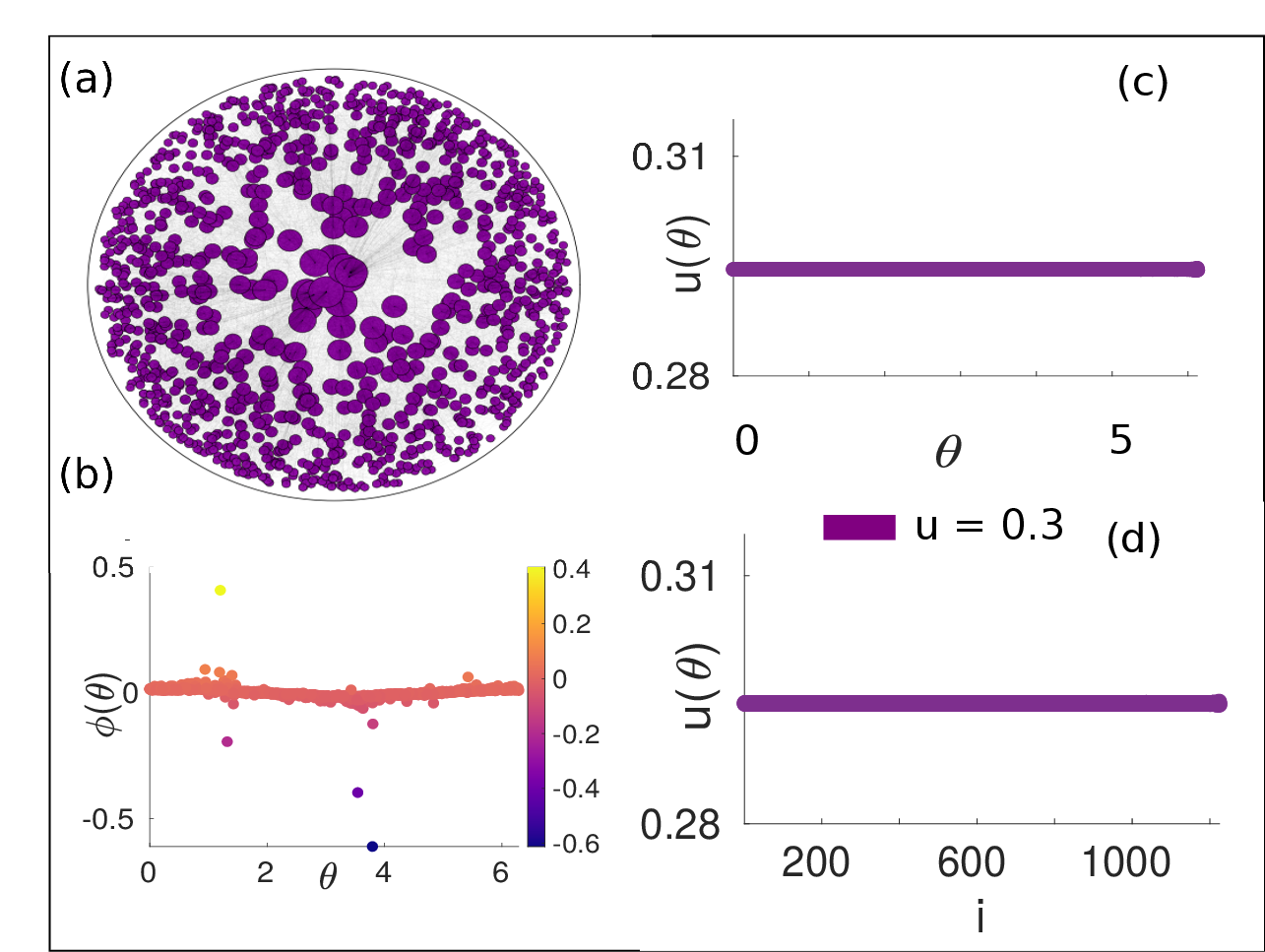}
\includegraphics[width=0.45\textwidth]{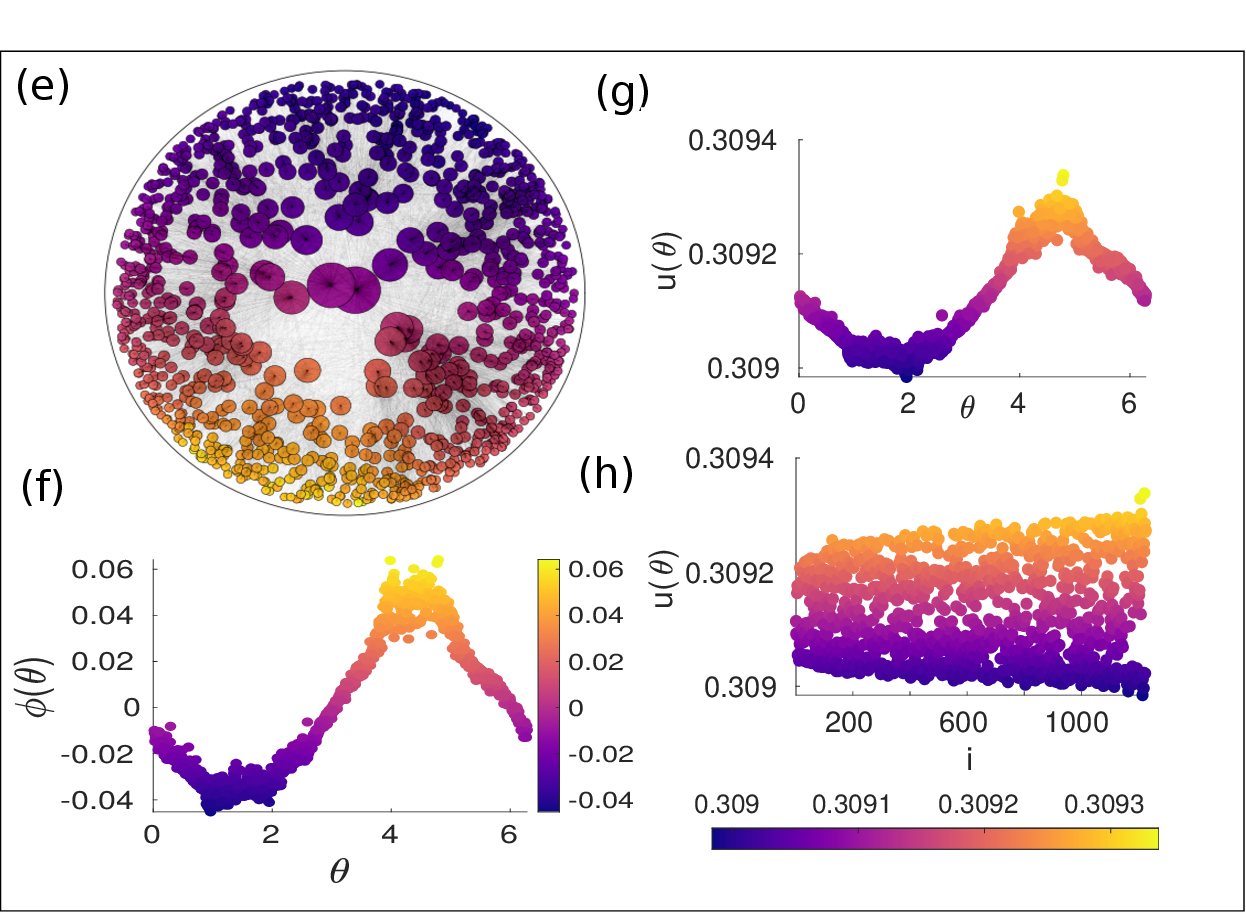}
    \caption{{Visualization in geometric networks; $N=1225$. ${\rm E}[k]=50$ (a)-(d) : $\beta=1.1$, $\Lambda) = 2.86$. (e)-(h):  $\beta=2.51$, $\Lambda=2.86$. Colors correspond to the density of $u$ in stationary state. Panels (a) and (e): nodes are located according to their co-ordinates in the hyperbolic space. Panels (b) and (f):  critical eigenvector plotted as a function of angular co-ordinate $\theta$. Panels (c) and (g): variation of $u$ as a function of $\theta$. Panels (d) and (h): variation of $u$ as a function of the nodes arranged in descending order of degree.} }
    \label{fig:hyperbolic_pattern}
\end{figure}
The critical eigenvectors for these cases are shown in Figs.\ref{fig:hyperbolic_pattern}(b) and (f) respectively, as a function of the angular coordinate $\theta$. The former shows a lack of any pattern as the values are the same for all $\theta$. In contrast, the latter shows a distinctive periodic structure, which has been shown to be indicative of Turing instability. These forms are evident in the values of ${\bf u}$ when plotted as a function of $\theta$; see Figs.\ref{fig:hyperbolic_pattern}(c) and (g).  The same are plotted as a function of the nodes when arranged in descending order of degree as shown in Figs.\ref{fig:hyperbolic_pattern}(d) and(h) and are seen to follow the qualitative patterns as observed for ER and scale free networks. 

\subsection{Eigenvector localization}
As has been discussed already, mathematically Turing instability is explained through Eq.\eqref{pert1} and for large time $(t\rightarrow \infty)$, the contributions are limited to only those terms for which $\Re(\gamma_s) >0$.
$\sigma$ can be adjusted such that it is slightly greater than $\sigma_c$ so that only one eigenmode contributes to  Eq.\eqref{pert1}. In such a case,
the corresponding ``pattern" has the same geometrical characteristics of the corresponding eigenvector. These unstable eigenmodes are therefore good indicators of the so called ``Turing patterns". For complex networks, as the nodes lack unique spatial measures or indicators,  patterns are difficult to visualize. Usually, the so called ``patterns" are visualized by arranging the nodes in descending orders of degree along an axis and plotting the corresponding values of the state variables.
In this section, the effect of clustering on the structure of the eigenvector is investigated, as the shape of the eigenvector governs the pattern formation. 

\subsubsection{$\mathbb{S}^1$ / $\mathbb{H}^2$ network}
Figure \ref{fig:hyperbolic_eigenvector1}shows the shape of the eigenvector $\phi$ as  function of nodes arranged in descending order of degree (top panel) and as a function of the angular coordinate $\theta$ (bottom panel) for a network with the following parameters  $(N, E[k], \beta, \Lambda) = (1225, 50, 1.1, 2.86)$. 
\begin{figure}[htbp]
\centering
\includegraphics[width=0.52\textwidth]{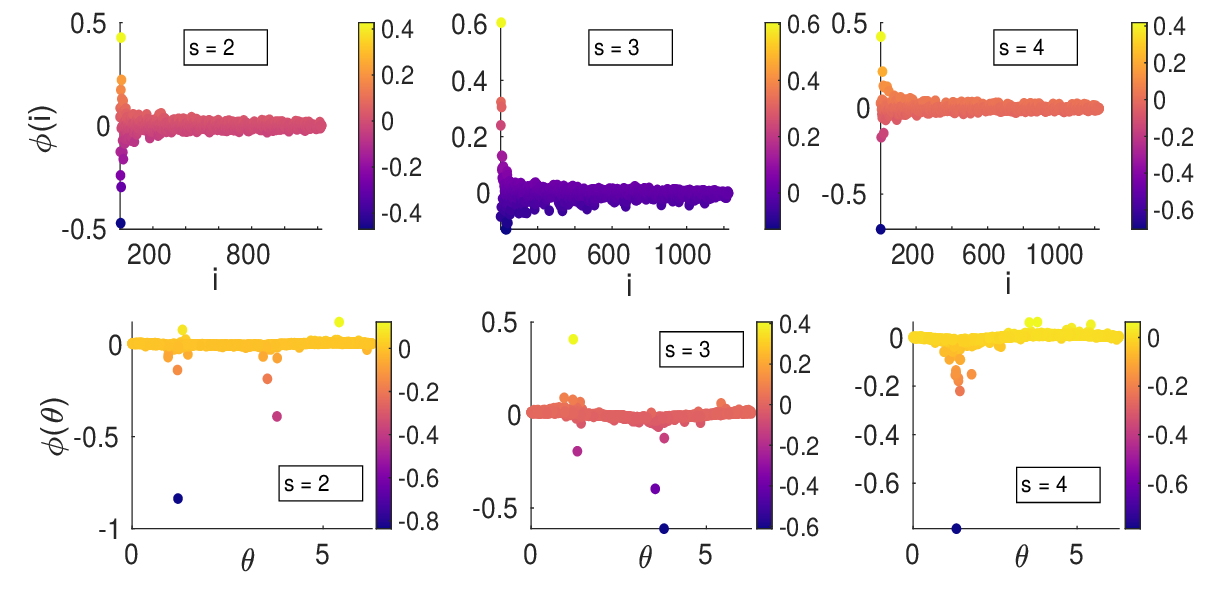}
    \caption{{Eigenvectors in geometric network: $N=1225$, $E[k]=50$, $\beta=1.1$, $\Lambda=2.86$; Top panel: variation of $\phi$ as a function of nodes arranged in descending order of degree; Bottom panel: variation of $\phi$ as a function of angular co-ordinate $\theta$. Both  panels show localization of $\phi$.}}
    \label{fig:hyperbolic_eigenvector1}
\end{figure}
Here, the eigenvectors $\phi_s$  for $s=2,3,4$ are shown. The corresponding eigenvalues are towards the right end of the spectrum and are close to zero.  It is observed that $\phi_s$ are predominantly clustered around zero, with very few values significantly greater than zero, indicating  the eigenvectors to be localized.
Figure \ref{fig:hyperbolic_eigenvector2} shows the shape of $\phi_s$  for $s=2,3,4$ but for a network with $\beta=2.51$ and the other parameters being the same as before.
\begin{figure}[t!]
\centering
\includegraphics[width=0.48\textwidth]{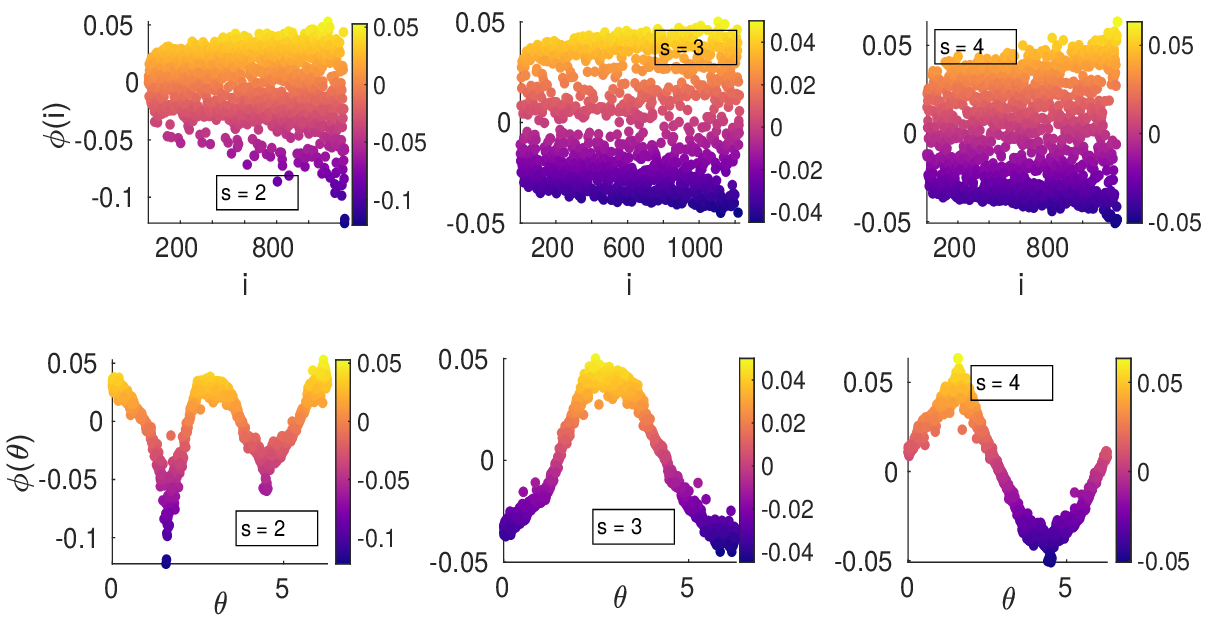}
    \caption{
    {Eigenvectors in geometric network: $N=1225$, $E[k]=50$, $\beta=2.51$, $\Lambda=2.86$. Top panel: variation of $\phi$ as a function of nodes arranged in descending order of degree. Bottom panel: variation of $\phi$ as a function of $\theta$; periodic behaviour is observed. } }
    \label{fig:hyperbolic_eigenvector2}
\end{figure}
This network is therefore highly clustered. The shape of the corresponding eigenvectors $\phi_s$ are significantly different from earlier; see Fig.\ref{fig:hyperbolic_eigenvector2}.
The top panel of Fig. \ref{fig:hyperbolic_eigenvector2} shows the variation of $\phi_s$ as a function of the nodes arranged in descending order of degree and unlike in the previous case, shows a wider variation about zero. A clear periodic structure is revealed when $\phi_s$ are shown as a function of $\theta$ (see bottom panel) indicating Turing instability \cite{van2023emergence} and illustrates the importance of clustering on Turing instability.
The extent of localization within an eigenvector is usually quantified using the inverse participation ratio (IPR)\cite{mcgraw2008laplacian,mimar2019turing}, which  for any eigenvector $\bf{\phi}$, is defined as 
\begin{align}
    IPR(\phi) = \frac{\sum_i \phi_i^4}{(\sum_i \phi_i^2)^2}.
\end{align}
IPR ranges from a minimum value of $1/N$ (for a normalized vector whose components are all of equal magnitude $1/\sqrt{N}$) to a maximum of 1 for a vector with only one non-zero component. The more localized is the vector, higher is the value of IPR. 
Figures \ref{fig:hyperbolic_ipr} show the plots of IPR for the two $\mathbb{S}^1$ / $\mathbb{H}^2$ networks considered. 
\begin{figure}[htbp]
\centering{
\includegraphics[width=0.23\textwidth]{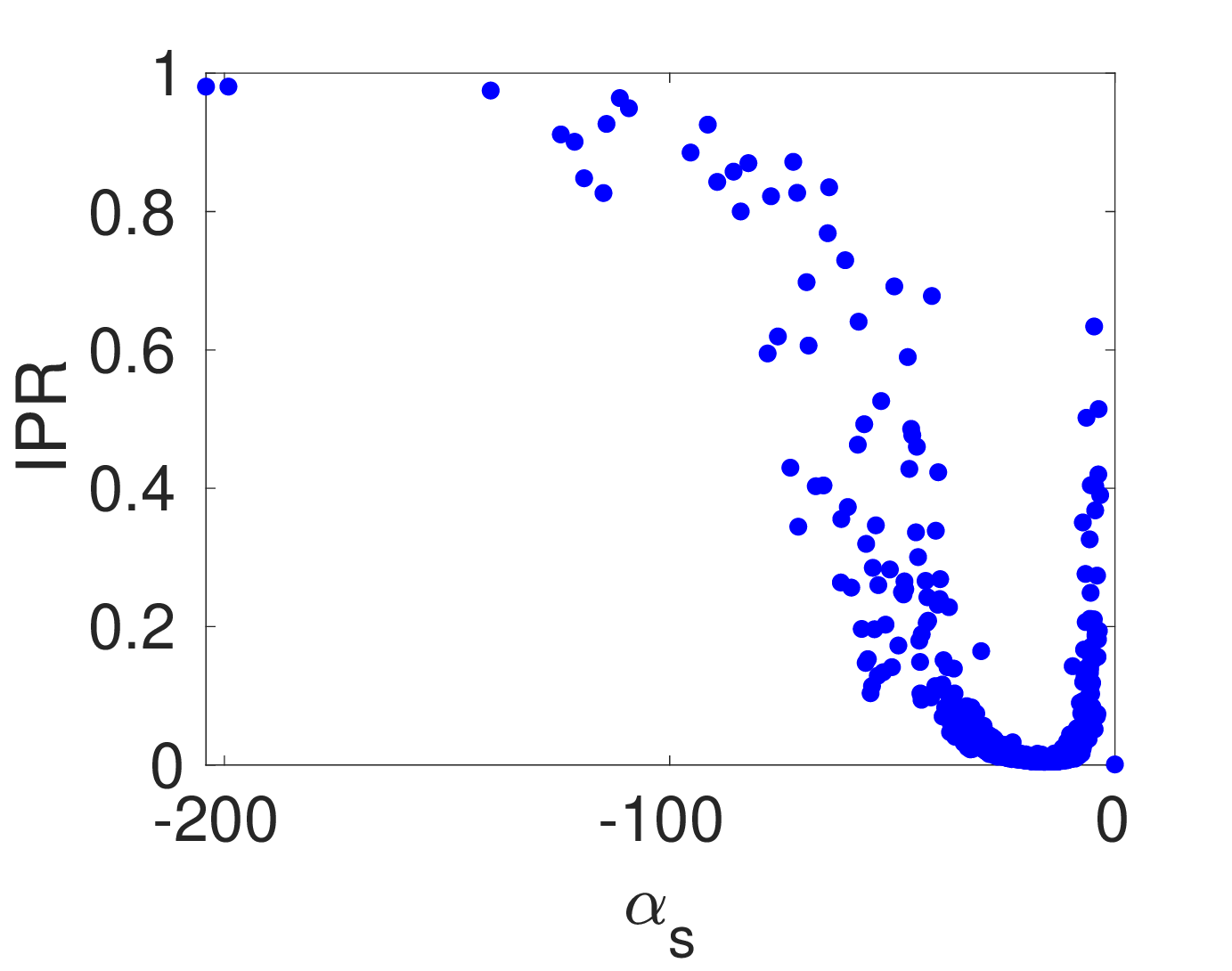}
\includegraphics[width=0.23\textwidth]{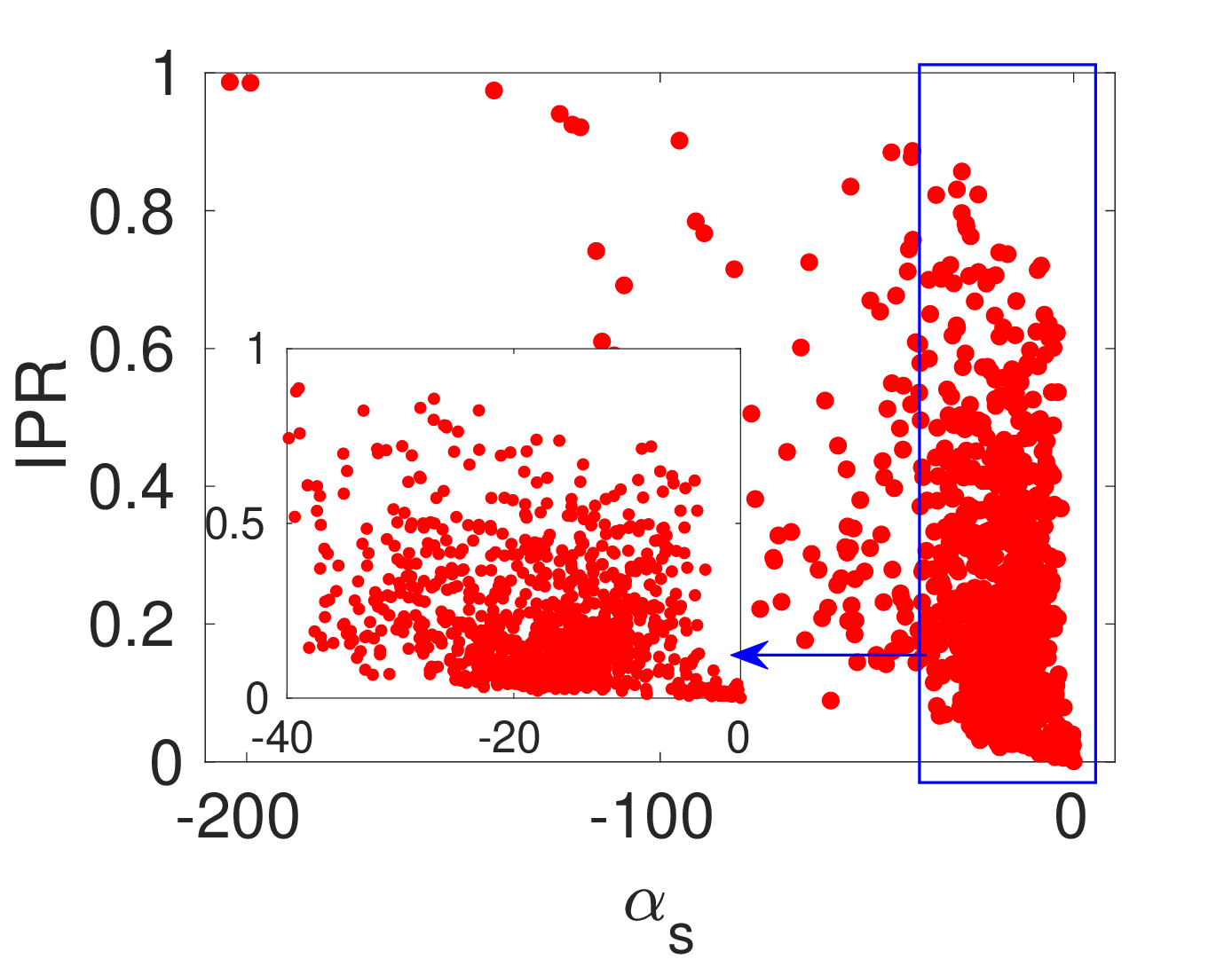}\\
\textbf{(a)} \hspace{1.9 in} \textbf{(b)}\\
\includegraphics[width=0.35\textwidth]{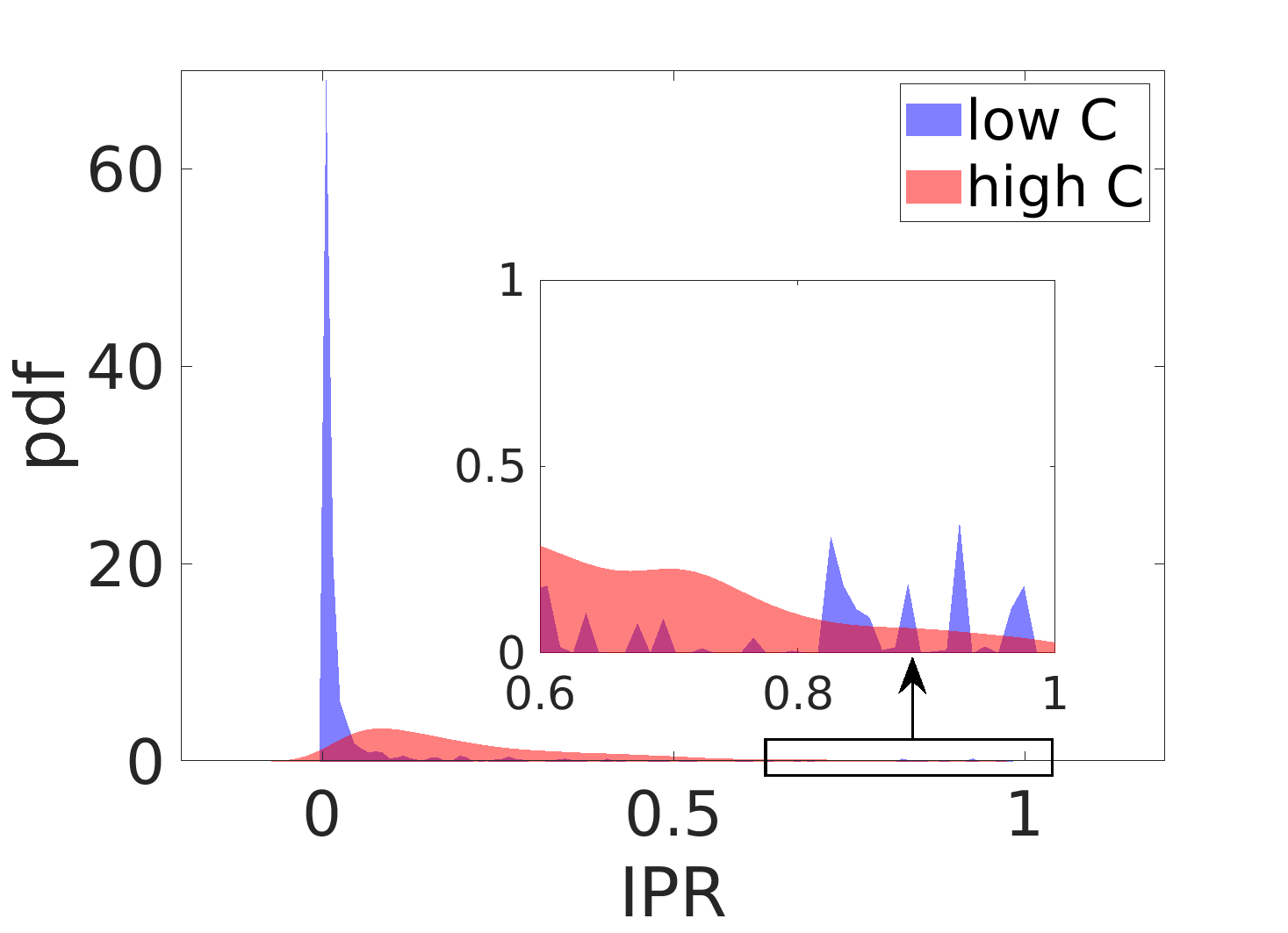}\\
\textbf{(c)}}
\noindent
    \caption{IPR versus $\alpha_s$ for $\mathbb{S}^1$ / $\mathbb{H}^2$ network; $N=1225$, ${\rm E}[k]=50$, $\lambda=2.86$ (a) weakly clustered $\beta= 1.1$, (b) highly clustered $\beta= 2.51$. Inset: zoomed view in the neighborhood of $\alpha_s=0$. (c) Probability density function of IPR values for  networks in (a)-(b); Inset: zoomed view of IPR near 1 .}
    \label{fig:hyperbolic_ipr}
\end{figure}
These figures reveal that (a) the eigenvectors corresponding to the extreme ends of the eigenspectrum tend to exhibit greater localization leading to higher IPR, and (b) increasing clustering in the network tends to increase the localization in the eigenvectors corresponding to the right end of the Laplacian eigenspectrum. In Figure \ref{fig:hyperbolic_ipr}(a), most of the IPR values are concentrated near zero, with higher IPR values appearing towards the tail, where the values approach 1. Conversely, in Figure \ref{fig:hyperbolic_ipr}(b), the eigenvalues near zero show no localization, as indicated by IPR values close to zero. However, as the eigenvalues move away from zero, there is a large group of eigenvectors that are moderately to strongly localized. Similarly, the eigenvalues near the tail end exhibit high localization. 
Figure \ref{fig:hyperbolic_ipr}(c) illustrates the pdf of IPR for $\beta=1.1$(low C) and $\beta=2.51$ (high C) represented by blue and red curves respectively. The abscissa represents IPR values, ranging from 0 to 1, with higher values indicating stronger localization of eigenvectors. For low C, there is a prominent peak near $IPR = 0$, suggesting that most eigenvectors are delocalized. In contrast, high C shows a lower peak near $IPR = 0$ and a broader distribution of higher IPR values, indicating a significant presence of localized eigenvectors. The inset plot highlights this difference by focusing on the higher IPR range (0.6 to 1), where high clustering has a more substantial presence of localized eigenvectors compared to low clustering. This demonstrates that increased clustering leads to greater localization of eigenvectors in the network.

\subsubsection{ER network}
\begin{figure}[t!]
\centering{
\includegraphics[width=0.5\textwidth]{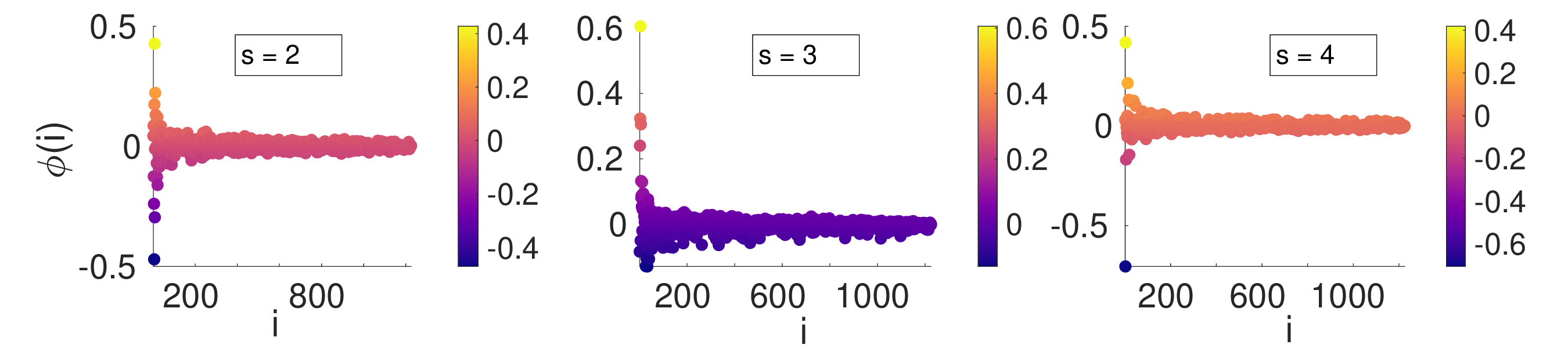}
\includegraphics[width=0.5\textwidth]{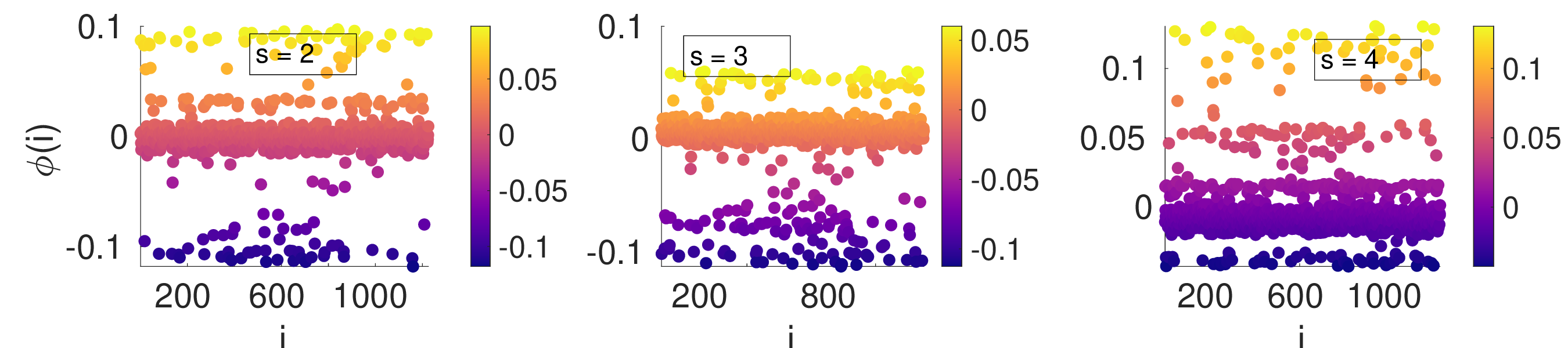}}
    \caption{{Eigenvectors of ${\bf L}$ of  ER network,  $E[k]=44$. abscissa: node indices arranged in descending order of degree; abscissa: eigenvector components (a) Top panel: $C=0.034$, $s=2,3,4$;  vectors are strongly localized (b) Bottom panel: $C=0.6$, $s=2,3,4$.  Eigenvector components fall near a small set of discrete values, giving the plot a striated appearance.} }
    \label{fig:ER_eigenvector}
\end{figure}
\begin{figure}[t!]
\centering{
\includegraphics[width=0.50\textwidth]{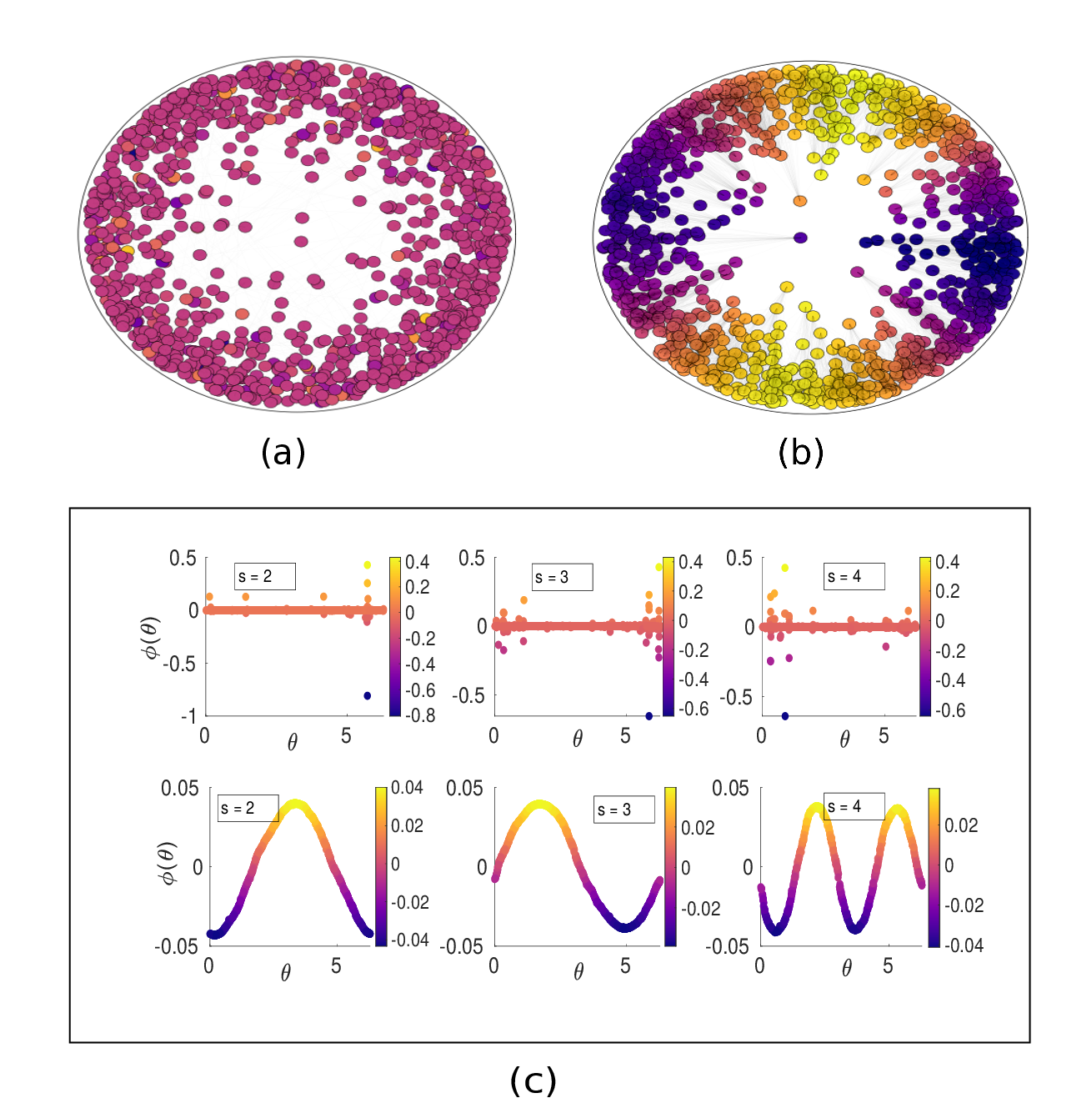}}
    \caption{{(a)-(b) Embedding of an ER network with $E[k]=44$ for $C=0.034$ and $C=0.6$ in the hyperbolic plane. The size of the nodes are proportional to the degree and the nodes are colored according to ${\boldsymbol \phi}_4$ values, (c) Top panel and bottom panel: variation of ${\boldsymbol \phi}_s$ for $s=2,3,4$ as a function of $\theta$ for low and high clustering respectively.  }}
    \label{fig:ER_embed}
\end{figure}
As discussed earlier in Fig.\ref{fig:eigenvector_critical}, the pattern formation in ER network depends on the structure of the critical eigenvector. In this section, the effect of clustering on the eigenvector structure is studied in ER networks. The top and bottom panels in Fig.\ref{fig:ER_eigenvector}  shows the eigenvectors ${\boldsymbol \phi}_s$ for $s=2,3,4$ for a low clustered ($C=0.034)$ and a high clustered ($C=0.6$) ER network with $E[k]=44$, respectively. In both the cases the nodes are arranged in a descending order of their degree. For the case of low clustering, it is observed that  ${\boldsymbol \phi}_s$ are clustered together and are close to zero with very few values being greater than zero implying the eigenvectors to be localized; similar behaviour were observed in the case of $\mathbb{S}^1$ / $\mathbb{H}^2$ network. For the same network but rewired such that $C=0.6$, the components of the vector ${\boldsymbol \phi}_2$, ${\boldsymbol \phi}_3$ and ${\boldsymbol \phi}_4$  are observed to be distributed predominantly along three regions (see Fig.\ref{fig:ER_eigenvector}) bottom panel).  The central cluster around zero is however more diffused than in the previous case. 

 ER networks being the limiting case of $\mathbb{S}^1$ / $\mathbb{H}^2$ network for $\Lambda \rightarrow \infty$, the effect of clustering on the structure of eigenvector can be better investigated with respect to the angular co-ordinate $\theta$ in the hyperbolic plane in which the network is embedded. This requires finding hidden degree $\kappa$ and angular co-ordinate $\theta$ for each node in the network. This is accomplished by using a tool named ``Mercator" \cite{garcia2019mercator}, which combines machine learning and maximal likelihood methods to find the optimal embedding consistent with $\mathbb{S}^1$ / $\mathbb{H}^2$ network. Embedding is performed for an the two ER networks with $E[k]=44$ and $C=0.034$ and $0.6$ respectively. The resulting network realizations are shown in Figs\ref{fig:ER_embed}(a)-(b) respectively. As is the usual practice in this visualization, the size of the nodes is taken to be proportional to its degree. As seen form the figure, all the nodes have comparable sizes which implies all the nodes have same hidden degree, that is, $\rho(\kappa) = \delta(\kappa - E[k])$ leading to a Poisson degree distribution with average degree $E[k]$ (in this case 44). Figure \ref{fig:ER_embed}(c) top panel shows the variation of eigenvectors ${\boldsymbol \phi}_s$ with respect to $\theta$ for $s=2,3,4$ for the low clustering embedded network. All the eigenvectors are observed to be localized about zero with some values significantly greater than zero. This can also be observed from Figure \ref{fig:ER_embed}(a) where the nodes are colored according to the values of ${\boldsymbol \phi}_4$. In the case of high clustered ER network, the eigenvectors corresponding to $s=2,3,4$ show well defined periodic structures with respect to angular co-ordinate $\theta$; see  bottom panel of Figure \ref{fig:ER_embed}(c). This periodic structures can also be observed from the network visualisation shown in Fig.\ref{fig:ER_embed}(b). This observation is consistent with that of the $\mathbb{S}^1$ / $\mathbb{H}^2$ networks, where eigenvectors corresponding to eigenvalues close to zero exhibit a well-defined periodic structure (see bottom panel of Fig.\ref{fig:hyperbolic_eigenvector2}). 
\begin{figure}[htbp]
\centering
\includegraphics[width=0.23\textwidth]{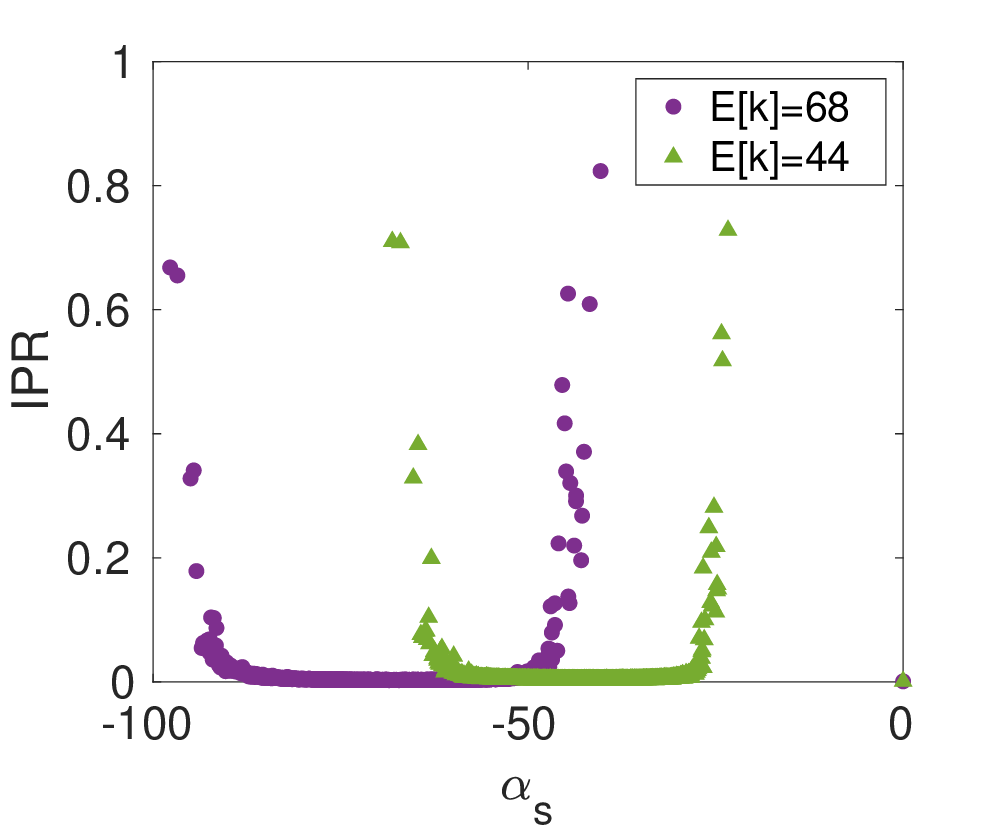}
\includegraphics[width=0.23\textwidth]{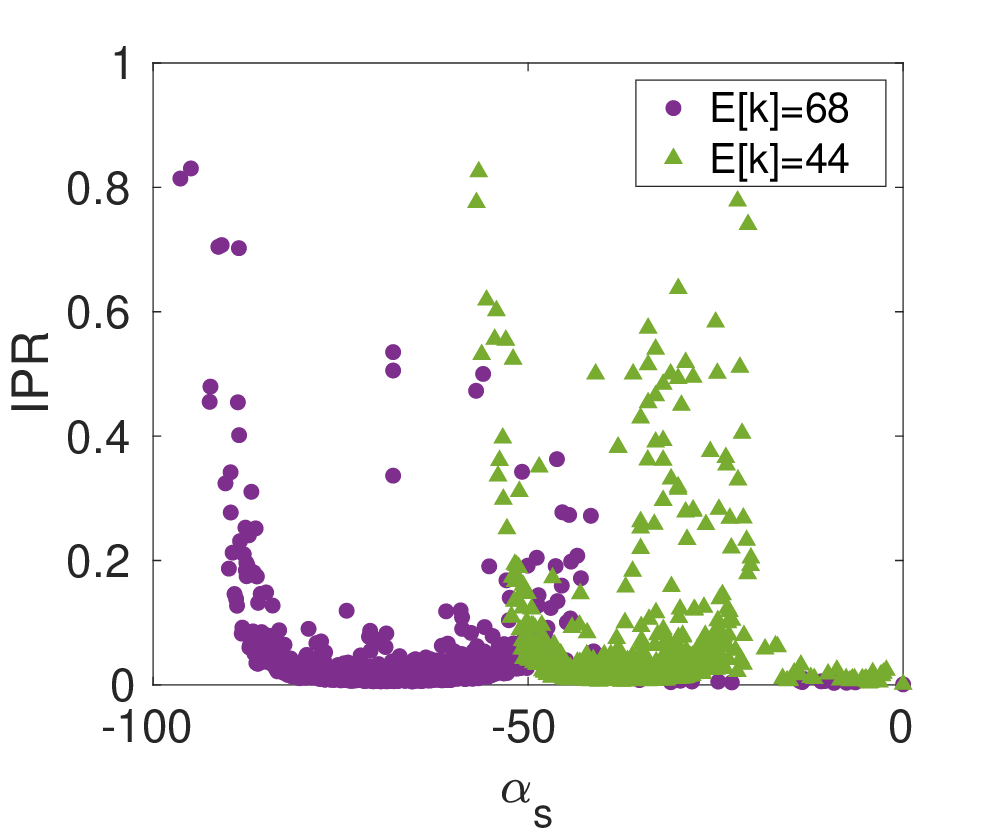}\\
\textbf{(a)} \hspace{1.8 in} \textbf{(b)}
    \caption{IPR of Laplacian eigenvectors plotted for ER networks with $E[k]=44,68$ for (a) Low clustering (b) High clustering}
    \label{fig:ipr_er}
\end{figure}
As in the case of $\mathbb{S}^1$ / $\mathbb{H}^2$ network, localization of eigenvectors is quantified using IPR. Figures \ref{fig:ipr_er}(a) and (b) shows the variation of IPR with $\alpha_s$ for ER networks with $E[k]=44,68$ for naturally occurring low clustering and $C=0.6$ respectively. For a network with low clustering, IPR values near the ends of the eigenspectrum are higher whereas the middle values are clustered near zero, indicating that the eigenvectors near the ends of the eigenspectrum are localized. Whereas, when the network is rewired such that the clustering increases, the eigenvalues close to zero exhibit no localization which results in $IPR=0$. This is also observed in the bottom panel of Fig. \ref{fig:ER_eigenvector} where the eigenvectors exhibit striated structure in contrast to the localized structure which results in IPR values close to 1 as seen in the top panel of Fig. \ref{fig:ER_eigenvector}.

\section{Conclusion} \label{conclusion}
Many real-world complex networks \cite{kunegis2013konect,ahn2006wiring,kim2014caenorhabditis,hagmann2008mapping} exhibit characteristic properties like degree heterogeneity, high clustering, small-worldness, and the presence of communities. Traditional models such as random and scale-free networks often fail to account for these properties adequately. However, the introduction of geometric soft configuration networks, which proved that the underlying metric space of complex networks as hyperbolic, has provided a more satisfactory explanation for these observed properties. It is noteworthy that random networks represent one of the limiting cases of geometric networks. While much research on Turing instability in complex networks has focused on tuning the average degree ($E[k]$) to trigger such instability, this approach overlooks the role of clustering—a prevalent feature in many complex networks. The current study addresses this gap by investigating the impact of clustering on Turing instability while maintaining a fixed degree distribution. The findings reveal that, even in scenarios where Turing instability may not be immediately apparent, tuning clustering can indeed induce it. This phenomenon is observed across random and scale-free networks, which serve as limiting cases of geometric networks. Furthermore, analysis of eigenvector localization properties uncovers distinct behaviors of eigenvectors as a result of tuning clustering in random and geometric networks, highlighting the role of clustering in shaping Turing patterns within complex networks.

\section*{Acknowledgments}
The authors would like to acknowledge the funding received from the Ministry of Education, Govt of India towards IoE Projects Phase II for the project titled Complex Systems \& Dynamics.

\section*{Author Declarations}
\subsection*{Conflict of Interest}
The authors have no conflicts of interest to declare.

\subsection*{Author Contributions}
{\bf SP}:Conceptualization (equal);  Formal analysis (lead); Investigation (lead); Methodology (equal);  Software (lead); Validation (equal); Visualization (lead); Writing – original draft (lead).
{\bf DJ}: Formal analysis (equal); Investigation (supporting); Methodology (supporting);  Software (supporting);  Writing – original draft (supporting).
{\bf SG}: Conceptualization (equal); Funding acquisition (lead); Investigation (supporting); Methodology (supporting); Project administration (lead); Supervision (lead); Visualization (supporting); Writing – review \& editing (lead).

\section*{Data Availability} No data has been used in this study.

\section*{References}
\bibliographystyle{ieeetr}
\bibliography{references}
\end{document}